%% file: main.tex
\documentclass[aps, pre, superscriptaddress,twocolumn,showpacs,longbibliography,amsmath,amssymb,nofootinbib]{revtex4-2}
\usepackage[utf8]{inputenc}
\usepackage[T1]{fontenc}
\usepackage{microtype}
\usepackage{graphicx}
\usepackage{mathbbol}
\usepackage{amssymb}
\usepackage{dsfont}
\usepackage{comment}

\usepackage[usenames,dvipsnames,svgnames,table]{xcolor}
\usepackage[colorlinks,linkcolor=blue,citecolor=blue,urlcolor=blue]{hyperref}
\usepackage{siunitx}
\usepackage{enumitem}

\usepackage{tikz}
\usetikzlibrary{positioning} %

\usepackage{booktabs}

\usepackage{etoolbox} %
\usepackage[english]{babel}
\usepackage{cleveref}
\crefrangelabelformat{equation}{(#3#1#4--#5#2#6)}
\crefname{equation}{Eq.}{Eqs.}
\Crefname{equation}{Equation}{Equations}

\makeatletter
\let\ORIbbl@fixname\bbl@fixname
\def\bbl@fixname#1{%
  \@ifundefined{languagealias@\expandafter\string#1}
    {\ORIbbl@fixname#1}
    {\edef\languagename{\@nameuse{languagealias@#1}}}%
}
\newcommand{\definelanguagealias}[2]{%
  \@namedef{languagealias@#1}{#2}%
}
\makeatother

\definelanguagealias{en}{english}

\renewcommand{\i}{\mathrm{i}}

\newcommand*\circled[1]{%
  \tikz[baseline=(char.base)]{
    \node[shape=circle,draw,inner sep=1pt] (char) {#1};}}

\def\equationautorefname~#1\null{Eq.~(#1)\null}

\def\equationautorefname~#1\null{Eq.~(#1)\null}

\begin{document}

    \title{Dynamical Phase Transitions Across Slow and Fast Regimes in a Two-Tone Driven Duffing Resonator}
	\date{\today}
 	\author{Soumya S. Kumar}
    \thanks{soumya.kumar@uni-konstanz.de}
	\affiliation{Department of Physics, University of Konstanz, 78464 Konstanz, Germany}

	\author{Javier del Pino}
    \affiliation{Department of Physics, University of Konstanz, 78464 Konstanz, Germany}
    \affiliation{Departamento de F\'{\i}sica Te\'{o}rica de la Materia Condensada and Condensed Matter Physics Center (IFIMAC), Universidad Aut\'{o}noma de Madrid, E-28049 Madrid, Spain}

    \author{Letizia Catalini}
    \affiliation{Laboratory for Solid State Physics, ETH Zürich, 8093 Zürich, Switzerland}
	\affiliation{Quantum Center, ETH Zürich, 8093 Zürich, Switzerland}
    \affiliation{Center for Nanophotonics, AMOLF, 1098XG Amsterdam, The Netherlands}

	\author{Alexander Eichler}
    \affiliation{Laboratory for Solid State Physics, ETH Zürich, 8093 Zürich, Switzerland}
	\affiliation{Quantum Center, ETH Zürich, 8093 Zürich, Switzerland}
	\author{Oded Zilberberg}
    \thanks{oded.zilberberg@uni-konstanz.de}
	\affiliation{Department of Physics, University of Konstanz, 78464 Konstanz, Germany}
	
	\DeclareGraphicsExtensions{.pdf,.png,.jpg}
	
\begin{abstract}
The response of nonlinear resonators to multifrequency driving reveals rich dynamics beyond conventional single-tone theory. We study a Duffing resonator under bichromatic excitation and identify a competition between the two drives, governed by their detuning and relative amplitudes. In the slow-beating regime, where the tones are closely spaced, the secondary tone acts as a modulation that induces dynamical phase transitions between coexisting stationary states. We introduce the cycle-averaged amplitude as an order parameter and map the resulting phase diagram as a function of the drive detuning and amplitude ratio, capturing the pronounced asymmetry observed for blue versus red detuning between the drives in experiment. We link the onset of these transitions to the resonance properties around the nonlinear stationary mode of the system. We apply a two-tone ansatz to this system, for the first time, to provide a comprehensive map of stationary states. Our results provide a framework for controlling driven nonlinear systems, enabling state manipulation, and sensing in nanomechanical, optical, and superconducting circuit platforms.
\end{abstract}

\maketitle	

\section{\label{sec:Introduction} Introduction}
    Nonlinear driven-dissipative systems, where coherent external drives compete with inherent dissipation and nonlinearity~\cite{dykman_fluctuating_2012, lifshitz_cross_review_nonlin}, are ideal for exploring diverse non-equilibrium phenomena relevant for climate physics~\cite{doi:10.1073/pnas.0705414105}, population dynamics~\cite{strogatz_nonlinear_2007, doi:10.1126/science.1219805}, nanotechnology~\cite{BUOT199373}, optics~\cite{Garmire:13, RevModPhys.86.1391}, and quantum technologies~\cite{harrington_engineered_2022}. Nonlinear dynamical behaviour such as bursting oscillations or relaxation-like cycles~\cite{vanderPol_relaxationoscillations}, characterised by alternating high-amplitude and low-amplitude activity, is observed in aerosol-cloud-precipitation systems~\cite{liu_double_2022}, circadian rhythms of various organisms~\cite{asgaritarghi_mathematical_2019}, and even optical fiber lasers subjected to weak optical injection~\cite{deshazer_bursting_2003}. 
    
    Multi-tone driving of nonlinear systems is ubiquitous: in Microelectromechanical (MEMS) and nanoelectromechanical (NEMS) resonators it enables the study of synchronisation, chaos, and is widely used in precision metrology and sensing~\cite{moser_ultrasensitive_2013, poggio_sensing_2013, leuch2016parametric, zhu_development_2019, wei_recent_2021, madiot2021bichromatic, stickler_zeptosensing_2023}; in ultra-cold atoms, it enables topological bands and Floquet-engineered states with controlled heating~\cite{sandholzer_floquet_2022, wang_topological_2023, chen_mitigating_2025}; it supports quantum memcapacitors in superconducting circuits~\cite{qiu_microwave_2024}; it generates magnonic frequency combs in magnomechanical systems~\cite{liu_generation_2023}; and it is a powerful tool to study dynamical phase transitions~\cite{yarmohammadi_dynamical_2025}. Such systems of slow, periodically forced resonators have been extensively studied in the context of mixed-mode oscillations~\cite{desroches_mixedmode_2012, desroches_bursting_minmodel_2019}. Even a simple two-tone driven Duffing resonator gives rise to complex dynamical behaviour which is not captured by existing methods~\cite{houri_pulse-width_2019, houri_generic_2020, madiot2021bichromatic, catalini_slow_2025, michaeli_optically_2025}. This complex behaviour, arising due to the interplay of dissipation, two drive tones, and nonlinearity, has led to sensing proposals~\cite{houri_pulse-width_2019, stickler_zeptosensing_2023}, realisation of chaos~\cite{houri_generic_2020} and even controlled transitions in multi-modal resonators~\cite{michaeli_optically_2025}. Despite this ubiquity, a systematic analysis that maps behaviours to parameter regimes is missing.

    Our understanding of nonlinear driven-dissipative systems is largely built on their response to single-tone driving. Traditional approaches to analyzing these systems rely on approximations that simplify the dynamics, such as the rotating-wave approximation or timescale separation, which neglect fast oscillating terms~\cite{Scully_Zubairy_1997}. While these approximations can be effective for single-frequency systems, they may not accurately capture the dynamics due to multiple tones. As such, established perturbative methods like Krylov-Bogoliubov~\cite{krylov_introduction_2016, nayfeh_nonlinear_1995}, Poincar\'{e}-Lindstedt~\cite{eichler_classical_2023} and others, while valuable for weakly-nonlinear single-tone systems or systems involving commensurate frequencies, are insufficient to capture dynamics arising due to incommensurate, low-detuned multi-tone drives where subdominant driving responses are usually treated as linear perturbations.

Recently, the complex trajectories of a two-tone driven Duffing resonator were examined through the lens of the changing phase-space flow~\cite{catalini_slow_2025}. A slow regime was identified where the system follows the flow to encircle coexisting attractors, in contrast to a fast regime where such global orbits are suppressed. While this perspective provides geometric intuition, it relies on a phenomenological reconstruction of the dynamics. A predictive framework that maps the boundaries of these dynamical phase transitions based on the resonator's response function is currently lacking.

     In this work, we develop an effective nonlinear response theory around driven stationary states to predict the thresholds for dynamical phase transitions. To this end, we first revisit the theory of linear and nonlinear response to establish the modified equations of motion. We then introduce an order parameter to map the dynamical phase diagram as a function of the detuning and relative strength of the secondary tone. Crucially, we extract an analytical threshold for the onset of these transitions, providing a vital tool for experimentalists to identify stable operating regimes. Furthermore, we apply a two-tone ansatz for the first time to this system, elucidating the complex dynamical responses beyond the reach of standard approximations. Our results leverage a linear-theory-based approach to gain new insight into the dynamics of a two-tone driven Duffing resonator. These results have implications across fields of physics in sensing and metrology, nonlinear optomechanics, optimized control of qubits, circuit quantum electrodynamics, and Floquet-engineering in ultra-cold atoms~\cite{houri_pulse-width_2019, bloch_param_coupled_2020, blais_cqed_2021, sandholzer_floquet_2022, valentin_bichromsemicondqubit_2024}. They are also important beyond physics in the study of tipping points in the context of early warning signs in climate dynamics, ecological systems, and socioeconomic models~\cite{strogatz_nonlinear_2007, scheffer2009early, KUEHN2011criticaltransitions, kuehn_warningsigns_2022}.

     The remainder of this paper is structured as follows: In Sec.~\ref{sec:System}, we introduce our model system: a Duffing resonator subject to a bichromatic drive. To build intuition, we first revisit the linear regime of the system in Sec.~\ref{sec:TwoLinResp}, establishing a rotating frame formalism that reveals how the two tones interact in the absence of nonlinearity. In Sec.~\ref{sec:OneNLresp}, we review the canonical single-tone driven Duffing resonator, detailing the harmonic balance method (HBM) used for its analysis. In Sec.~\ref{sec:TwoNLresp}, we analyze the slow-beating regime, treating the secondary tone as a slow modulation that can induce dynamical phase transitions, and we introduce an order parameter to map out the resulting phase diagram. In Sec.~\ref{sec:Phase Transition Boundaries}, we refine a linear-response-based approach with nonlinear corrections to analytically approximate the observed transition boundaries. In Sec.~\ref{sec:Extension of Ansatz}, we apply an extended (two-tone) ansatz to this system for the first time to provide a comprehensive map of stationary states and connect the results with the analytical boundaries in Sec.~\ref{sec:Phase Transition Boundaries}. Finally, we conclude in Sec.~\ref{sec:Conclusion/Outlook} with a summary of our key findings and an outlook on future research directions.

\section{\label{sec:System}System}
    
    The dynamics of a nonlinear Duffing (Kerr) resonator, subject to a bichromatic (two-tone) drive, is described by the equation of motion
    \begin{equation}\label{eq:DuffingEOMTwoTone}
        \ddot{x}+\Omega_0^2 x+\Gamma \dot{x}  + \alpha x^3 = \sum_{i=1,2}\mathrm{Re}[F_i e^{i(\Omega_i t + \theta_i)}]\,.
    \end{equation}
   Here, $x$ denotes the resonator displacement, $\Omega_0$ is the natural resonance frequency, $\Gamma$ represents the damping coefficient, and $\alpha$ is the Duffing nonlinearity. We work in units where the resonator mass is $m=1$. The system is driven by two tones with amplitudes $F_i$, frequencies $\Omega_i$, and phases $\theta_i$, where $i=1,2$.

    The sign of the Duffing coefficient $\alpha$ dictates the nature of the nonlinearity: $\alpha < 0$ ($\alpha > 0$) corresponds to a softening (hardening) spring characteristic with increasing displacement. This sign also governs qualitative changes in the potential energy landscape of the resonator, see Fig.~\ref{fig:fig1}(a). Our analysis considers the weak nonlinearity regime, where the quartic term remains a perturbation to the harmonic potential. Furthermore, we assume a negative Duffing nonlinearity $\alpha<0$; a similar analysis for $\alpha>0$ will yield qualitatively similar results under proper tuning of parameters. 

\section{\label{sec:TwoLinResp}Two-tone linear response theory}
    We first revisit the linear response of a resonator driven by two tones. The equation of motion of a linear damped resonator [$\alpha=0$ in Eq.~\eqref{eq:DuffingEOMTwoTone}] is readily solved in Fourier space~\cite{Fourier} %
    , where the resulting Fourier amplitude of the system reads
    \begin{align}\label{eq:SHOResponseLab}
        \tilde{x}(\omega)&=\sum_{i=1,2}{F_{i}}\chi_{\raisebox{-0.3ex}{$\scriptscriptstyle \Omega_0$}}(\omega)\delta(\Omega_i-\omega)\,,
    \end{align}
    with $\chi_{\raisebox{-0.3ex}{$\scriptscriptstyle \Omega_0$}}(\omega)=\left(\Omega_{0}^2-\omega^2-i\Gamma\omega\right)^{-1}$ the susceptibility of the damped harmonic resonator~\cite{arfken_mathematical_2013}, which characterizes the resonator's inherent frequency-dependent response, see Fig.~\ref{fig:fig1}(b).    
    In linear systems, the equation of motion precludes frequency mixing. Consequently, the total response~\eqref{eq:SHOResponseLab} is a superposition of individual responses. The response amplitude (and thus the power) depends on the drive strength $F_i$, the resonator susceptibility $\chi_{\raisebox{-0.3ex}{$\scriptscriptstyle \Omega_0$}}(\omega)$, and the detuning; see Fig.~\ref{fig:fig1}(b). The response diminishes as the drive frequency deviates from $\Omega_0$. Hence, the combined effects of its strength and detuning determine which drive leads to a stronger response.

\newlength{\figwidth}
\setlength{\figwidth}{0.9\columnwidth}
\newlength{\figheight}
\setlength{\figheight}{0.675\columnwidth}

\newcommand{\fontselect}{\normalsize\bf} %

\tikzset{
  labelNode/.style={anchor=south east, above left=0cm and 0cm of anchor, font=\fontselect}, %
  redAnchorNode/.style={anchor=center, left=-0.1cm of anchor, circle, fill=red, minimum size=0.2cm}, %
  graphicNode/.style={anchor=north west, below right=-0.325cm and -0.105cm of anchor}, %
}

\newtoggle{draft}
	\togglefalse{draft} %
\begin{figure}

    \begin{tikzpicture}[every node/.style={inner sep=0pt}]
    
    \iftoggle{draft}{\node[anchor=south west, rectangle, draw, red, line width=2pt, minimum width=\figwidth, minimum height=\figheight] at (0,0) {};}
    {\node[anchor=south west, rectangle, fill=white, minimum width=\figwidth, minimum height=\figheight] at (0,0) {};};

    \node (anchor) at (-0.685,1\figheight) {}; %
    \node[graphicNode] {\includegraphics[width=0.495\columnwidth]{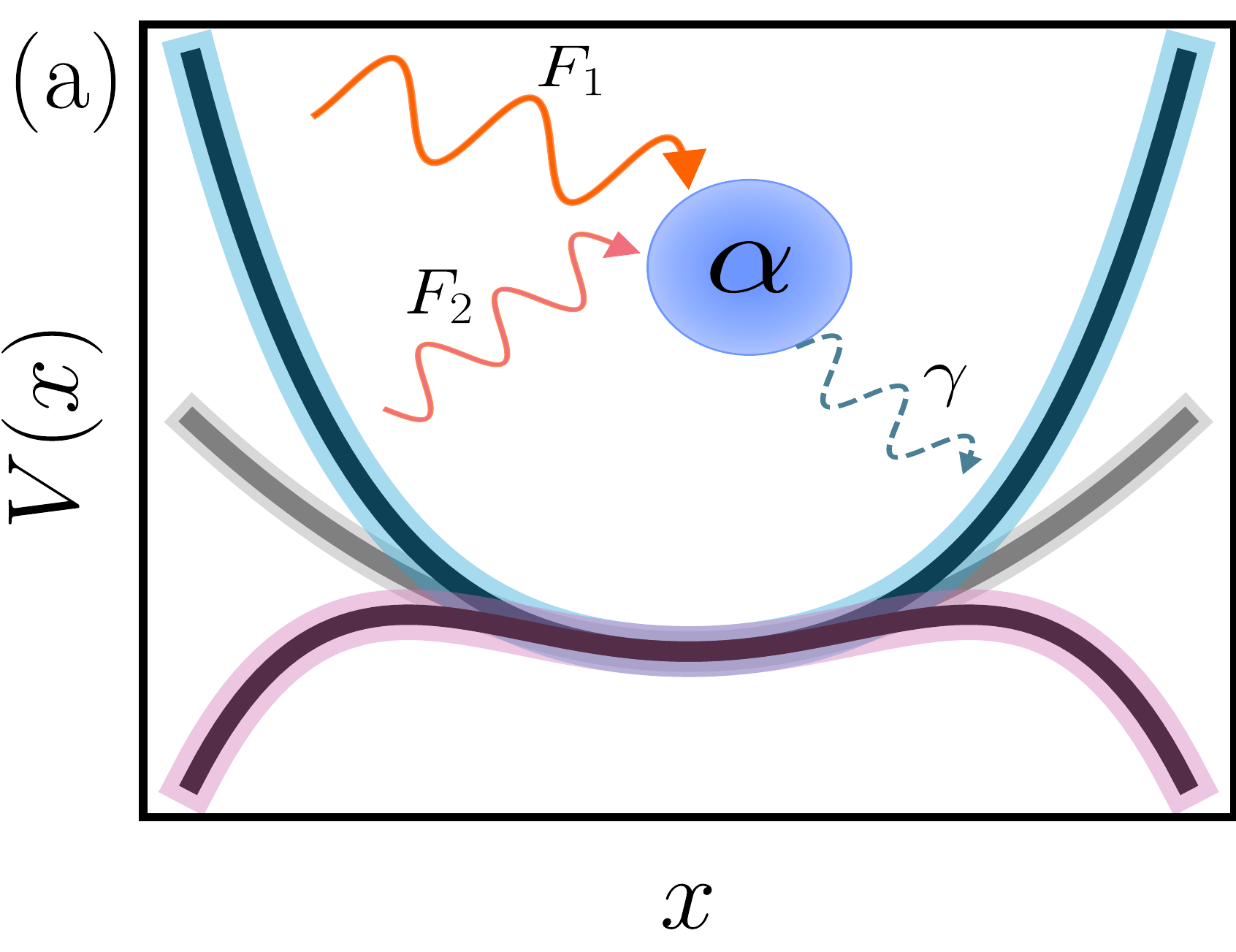}}; %
    \iftoggle{draft}{\node [redAnchorNode] {};}; %

    \node (anchor) at (-0.685,0.415\figheight) {}; %
    \node[graphicNode] {\includegraphics[]{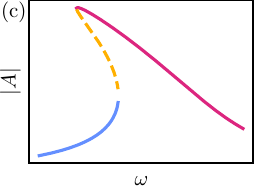}}; %
    \iftoggle{draft}{\node [redAnchorNode] {};}; %
    
    \node (anchor) at (0.425\columnwidth,0.989\figheight) {}; %
    \node[graphicNode] {\includegraphics[]{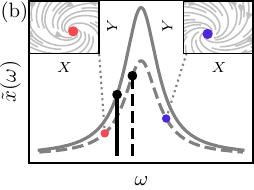}}; %
    \iftoggle{draft}{\node [redAnchorNode] {};}; %
    
    \node (anchor) at (0.425\columnwidth,0.415\figheight) {}; %
    \node[graphicNode] {\includegraphics[]{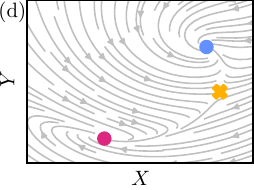}}; %
    \iftoggle{draft}{\node [redAnchorNode] {};}; %
    
    \end{tikzpicture}
    \caption{The Duffing resonator's linear response to two drives and nonlinear response to a single-drive. (a) Potential energy landscape $V(x)$ of a Duffing resonator [cf.~Eq.~\eqref{eq:DuffingEOMTwoTone}] with positive (blue), vanishing (grey), and negative (red) nonlinearity $\alpha$. Inset: Sketch of a Duffing resonator subject to two different drives (wavy inbound arrows) and dissipation (dashed outbound arrow).~(b) Stationary amplitude $\tilde{x}(\omega)$ of a linear resonator driven by two tones [cf.~Eq.~\eqref{eq:SHOResponseLab}]. The solid and dashed curves represent the scaled susceptibility ${F_{i}}\chi_{\raisebox{-0.3ex}{$\scriptscriptstyle \Omega_0$}}(\omega)$ for different drive strengths $F_{i}$. Note that a weaker drive closer to resonance (dashed) can yield a higher response than a stronger detuned drive (solid). Insets: Rotating frame vector flow [cf.~Eq.~\eqref{eq:Linear_harmonic_equations}]; a positively detuned drive induces clockwise motion, while a negatively detuned one induces anti-clockwise motion.~(c) Single-tone nonlinear response [cf.~Eq.~\eqref{eq:1toneSlowFlowEqDuffing}] exhibiting bistability between low (blue) and high (pink) amplitude branches, separated by an unstable saddle (yellow dashed). We use $\Omega_{0}=2\pi$, $\Gamma=6.2\times10^{-4}$, $F_{1}=4.81\times10^{-4}$.~(d) Phase space vector flow (grey lines) for a drive fixed in the bistable region ($\Delta_{10}=-9.02\times10^{-4}$), showing two counter-rotating attractors (pink and blue dots) separated by a saddle (yellow cross).}
    \label{fig:fig1}
\end{figure}

    To further analyze the system's response to the two driving tones, we first assume that the drive $F_{1}$  (at frequency $\Omega_{1}$) elicits the dominant response. Accordingly, we reformulate the problem in a frame co-rotating with this main drive component. To this end, the resonator's displacement $x(t)$ is expressed via the transformation 
    \begin{equation}\label{eq:rotating_quadratures}
        x(t)=X(t)\cos{\Omega_1 t}-Y(t)\sin{\Omega_1 t}\,,
    \end{equation}
    where $X(t)$ and $Y(t)$ are the time-dependent resonator quadratures. We decompose the overall drive $F(t)$ from Eq.~\eqref{eq:DuffingEOMTwoTone} relative to $\Omega_{1}$ as
    \begin{equation}\label{eq:ForceAfterFreqDecomposition}
        F(t) =\mathrm{Re}\left[e^{i \left(\Omega_{1}t+\theta_{1}\right)}\left(1 + he^{i\left(\Delta_{21}t+\theta_{2}-\theta_{1}\right)}\right)\right] F_1\,,
    \end{equation}
    where $\Delta_{21}=\Omega_2-\Omega_1$ is  the detuning between the drive frequencies, $h=\frac{F_{2}}{F_{1}}$ is the relative strength of the two tones.

    Substituting the transformation~\eqref{eq:rotating_quadratures} into the linear equation of motion [Eq.~\eqref{eq:DuffingEOMTwoTone} with $\alpha=0$], we obtain the coupled equations for the quadratures by equating the coefficients of the $\sin(\Omega_{1}t)$ and $\cos(\Omega_{1}t)$ contributions independently:
    \begin{align}\label{eq:Linear_harmonic_equations}
        \begin{bmatrix}
    \ddot{X} \\
    \ddot{Y}
    \end{bmatrix} \!+\!
     \begin{bmatrix}
    \Gamma & 2\Omega_1 \\
    -2\Omega_1 & \Gamma
    \end{bmatrix}  \!
    \begin{bmatrix}
    \dot{X} \\
    \dot{Y}
    \end{bmatrix}
    \!+\!
    \begin{bmatrix}
    \tilde\Delta_{01}^2 & \Gamma \Omega_1 \\
    -\Gamma \Omega_1 & \tilde\Delta_{01}^2
    \end{bmatrix} \!
    \begin{bmatrix}
    X \\
    Y
    \end{bmatrix}
    = \vec{F}\, ,
    \end{align}
    with effective force $\vec{F}=F_1[1 + h \cos (\Delta_{21} t),- h \sin (\Delta_{21} t)]^T$  written in the $\Omega_{1}$-frame, with $\tilde\Delta_{01}=\sqrt{\Omega_{0}^{2}-\Omega_{1}^{2}}$.
These \textit{harmonic equations} for the quadratures can be used to visualize the vector flow in a $2D$ rotating phase space, as is measured in experiments with a lock-in amplifier~\cite{mukamel_principles_1995}.
{Closed-loop stationary orbits in the phase space of the laboratory frame appear as stable states, i.e., fixed-point attractors, in the rotating phase space spanned by the quadratures $X(t)$ and $Y(t)$. Additionally, a negatively detuned drive results in an anticlockwise vector flow around the attractor, whereas a positively detuned drive leads to a clockwise vector flow, see insets in Fig.~\ref{fig:fig1}(b), lending a notion of \textit{chirality} to the attractor in the rotating frame~\cite{soriente_distinctive_2021, heugel_role_2023, dumont_energy_2024,villa2025topological}}.
  
The dynamics of the quadratures, governed by Eq.~\eqref{eq:Linear_harmonic_equations}, directly determine the time-dependent squared amplitude of the resonator's response in the rotating frame, $A^2_{\raisebox{-0.3ex}{$\scriptscriptstyle \Omega_0$}}(t, h) = X^2 (t, h) + Y^2 (t, h)$, which quantifies the power measured with a lock-in amplifier at $\Omega_1$. To find $A^2_{\raisebox{-0.3ex}{$\scriptscriptstyle \Omega_0$}}(t, h)$, we first solve Eq.~\eqref{eq:Linear_harmonic_equations} for $\tilde{X}(\omega)$ and $\tilde{Y}(\omega)$ via Fourier transformation. A further inverse Fourier transform yields (cf.~Appendix~\ref{sec:derivation_linearpower_modulated} for details):
\begin{multline}\label{eq:linearPoweratPumptimedependence}
    A^2_{\raisebox{-0.3ex}{$\scriptscriptstyle \Omega_0$}}(t, h) = F_1^2(1|\chi_{1}|^2\!+\!h^2|\chi_{2}|^2\!)+\!2F_1^2h|\chi_{1}|^2|\chi_{2}|^2\! \times \\
    \biggl[a_{\raisebox{-0.3ex}{$\scriptscriptstyle \Omega_0$}} \cos\left(\Delta_{21}t\right)
    + b_{\raisebox{-0.3ex}{$\scriptscriptstyle \Omega_0$}}\sin\left(\Delta_{21}t\right)  \biggr] \, ,
\end{multline}
where $\chi_i\equiv \chi_{\raisebox{-0.3ex}{$\scriptscriptstyle \Omega_0$}} (\omega=\Omega_i)$ is the resonator susceptibility at the drive frequency $\Omega_i$ and
\begin{align}
a_{\raisebox{-0.3ex}{$\scriptscriptstyle \Omega_0$}}&=|\chi_{1}|^{2}\!+\!\Delta_{21}\!\left[\Gamma^{2}\Omega_{1}\!+\!(\Delta_{21}\!+\!2\Omega_{1})\left(\Omega_{1}^{2}\!-\!\Omega_{0}^{2}\right)\right],\label{eq:shorthand_a}\\
b_{\raisebox{-0.3ex}{$\scriptscriptstyle \Omega_0$}}&=\Gamma\Delta_{21}\!\left[\Omega_{1}(\Delta_{21}\!+\!\Omega_{1})\!+\!\Omega_{0}^{2}\right]\,.\label{eq:shorthand_b}
\end{align}
Crucially, Eq.~\eqref{eq:linearPoweratPumptimedependence} reveals that although the linear resonator responds to each drive tone independently, the power associated with the $\Omega_1$ component of motion is not simply the direct response $F_1^2|\chi_1|^2$. Instead, $A^2_{\raisebox{-0.3ex}{$\scriptscriptstyle \Omega_0$}}(t, h)$ incorporates two additional contributions from the secondary tone ($F_2,\, \Omega_2$): (i) a static term $F_2^2|\chi_2|^2$, and (ii) a time-dependent cross-term proportional to $F_1 F_2$, originating from the beating between the two drive frequencies. This cross-term signifies a temporal modulation of the primary response amplitude by the secondary tone.

\section{\label{sec:OneNLresp}Nonlinear single-tone response}  
    While the linear response to two tones already exhibits complex temporal dynamics [cf.~Eq.~\eqref{eq:linearPoweratPumptimedependence}], the introduction of the nonlinearity fundamentally alters the system's behaviour. For a Duffing resonator driven by a single drive [$F_2=0$ in Eq.~\eqref{eq:DuffingEOMTwoTone}], the nonlinear cubic term induces frequency mixing, i.e., the generation of new frequency components such as harmonics 
    of the input frequency~\cite{eichler_classical_2023}. This precludes an exact analytical solution via Fourier transformation. However, approximate analytical solutions can be obtained via perturbative methods such as the Krylov-Bogoliubov method~\cite{nayfeh_nonlinear_1995, krylov_introduction_2016}, the Poincar\'{e}-Lindstedt method~\cite{nayfeh_perturbation_2000}, secular perturbation theory~\cite{lifshitz_cross_review_nonlin, rand_richard_lecture_2012}, or the HBM~\cite{krack_harmonic_2019}. These methods rely on separating the timescale dominating the stationary system response from the faster timescales determined by the resonance frequency, or the drive acting on the system.    

  In this section, we use the HBM which involves (i) switching to the rotating quadratures via Eq.~\eqref{eq:rotating_quadratures}, (ii) assuming the quadratures evolve on a timescale $T$ much larger than the system oscillations $\left(T\gg\frac{2\pi}{\Omega_1}\right)$ [thereby rendering $X(t)\rightarrow X(T), Y(t) \rightarrow Y(T)$], such that their variation over a single drive cycle ($2\pi/\Omega_1$) is small, and finally (iii) ``balancing harmonics'', i.e., matching terms with harmonics rotating at the same frequency $\Omega_{1}$~\cite{kosata_harmonicbalancejl_2022}. This slow-evolution ansatz (integral to many perturbative approaches) justifies neglecting higher-order time derivatives of the quadratures (i.e., $\ddot{X}, \ddot{Y} \approx 0$). The resulting autonomous system of two coupled first-order differential equations reads
    \begin{align}\label{eq:1toneSlowFlowEqDuffing}
    \begin{bmatrix}
    \dot{X} \\
    \dot{Y}
    \end{bmatrix}
    =
    \begin{bmatrix}
    -\frac{\Gamma}{2} & \frac{3 \alpha {A}^2 + 4 \tilde{\Delta}_{01}^2}{8 \Omega_1} \\
    -\frac{3 \alpha {A}^2 + 4 \tilde{\Delta}_{01}^2}{8 \Omega_1} & -\frac{\Gamma}{2}
    \end{bmatrix}
    \begin{bmatrix}
    X \\
    Y
    \end{bmatrix}
    +
    \frac{\vec{F}_1}{2\Omega_1}\,,
    \end{align} 
    in terms of the response amplitude, ${A}=\sqrt{X^2+Y^2}$, at frequency $\Omega_1$, with $\vec{F}_1=F_1[ \sin{\theta_1},  \cos{\theta_1}]^T$. 
    
    The stationary states for the single-tone driven Duffing resonator are found by setting $\dot{X}=\dot{Y}=0$ in Eq.~\eqref{eq:1toneSlowFlowEqDuffing}. The drive strength $F_1$ and frequency $\Omega_1$ determines the system's stationary characteristics, including the response amplitude and the location of bifurcation points. At low driving strengths, the Duffing exhibits a quasi-linear response, resembling a tilted Lorentzian. Increasing $F_{1}$ causes the response to bifurcate, leading to a high-amplitude branch and a low-amplitude branch, where a branch is a continuum of stationary states as a function of a system parameter. This bifurcation underlies the characteristic ``shark-fin'' profile observed when plotting the response amplitude against the drive frequency, see Fig.~\ref{fig:fig1}(c). The dynamics governed by Eq.~\eqref{eq:1toneSlowFlowEqDuffing} can be visualized as a 2D vector flow in phase space, see Fig.~\ref{fig:fig1}(d). 
   When the system is in the bistable region, there are two attractors in phase space with opposite chirality: 1 (-1) for the clockwise (anti-clockwise) high (low) amplitude state~\cite{villa2025topological}.
    
    We find the frequency range in which bistability occurs by analyzing the number of real roots of the response amplitude equation~\cite{papariello_ultrasensitive_2016}, which is accomplished by combining the two stationary conditions in Eq.~\eqref{eq:1toneSlowFlowEqDuffing} into
    \begin{equation}\label{eq:AmplitudePolynomial}
        A^2\left( {\Gamma^2} + \left( \frac{3\alpha A^2 }{4\Omega_1} + \frac{\Omega_1^2 - \Omega_0^2}{\Omega_1} \right)^2 \right) = \frac{F_1^2}{\Omega_1^2}\,.
    \end{equation}
    This equation is a cubic polynomial in $A^2$. The number of its real solutions, corresponding to the stable and unstable states, changes when its discriminant vanishes. This occurs at two distinct saddle-node bifurcations, where the low- and high-amplitude branches lose stability.
    This vanishing discriminant condition reads
    \begin{multline}\label{eq:duffing_bif}
|F_{\mathrm{lb/hb}}|^{2}\!=\!\frac{8}{81\alpha}\!\Big[\!\left(\tilde\Delta_{01}^{4}-3\Gamma^{2}\Omega_{1}^{2}\right)^{\frac{3}{2}}\! \\ \pm\tilde\Delta_{01}^{2}\left(\tilde\Delta_{01}^{4}+9\Gamma^{2}\Omega_{1}^{2}\right)\!\Big]\,,
    \end{multline}
    and establishes the relationship between the critical drive strength $|F_{\mathrm{lb/hb}}|^2$ for the low/high amplitude branch to bifurcate, and the bifurcation frequencies $\Omega_1$, where the instability occurs. 
    Thus, we can explicitly define the drive strength required to induce a bifurcation at a given frequency, or conversely, the frequencies at which bifurcations occur for a fixed drive strength. Crucially, for fixed ($\Delta_{01},\Gamma,\Omega_1,\alpha$) these thresholds are not equal: to jump up (low amplitude $\to$ high amplitude) the drive amplitude, $F_1$, must exceed the $F_{\mathrm{lb}}$, while to jump down (high $\to$ low) it must be reduced below $F_{\mathrm{hb}}$. Thus, sweeping the drive amplitude traces a hysteresis loop with direction-dependent switching.

\section{\label{sec:TwoNLresp}Two-tone nonlinear response}

    We now address the nonlinear Duffing resonator subject to a two-tone drive.
    A fundamental challenge in the two-tone driven Duffing resonator is that transforming to a co-rotating frame at a single drive frequency does not yield an autonomous system of equations. 
    This inherent non-autonomy was already evident in the linear case [cf.~Eqs.~\eqref{eq:ForceAfterFreqDecomposition} and~\eqref{eq:linearPoweratPumptimedependence}]; it occurs because a single rotating frame cannot simultaneously eliminate the time dependence from both drive frequencies. This complicates the determination of stationary states, as the problem cannot be directly reduced to solving a time-independent algebraic system for fixed ansatz amplitudes, in contrast to the single-tone scenario, cf.~Eq.~\eqref{eq:1toneSlowFlowEqDuffing}.

\newlength{\figheightsmall}
\setlength{\figheightsmall}{0.39\columnwidth}

\begin{figure}
    \centering
    \includegraphics[width=\columnwidth]{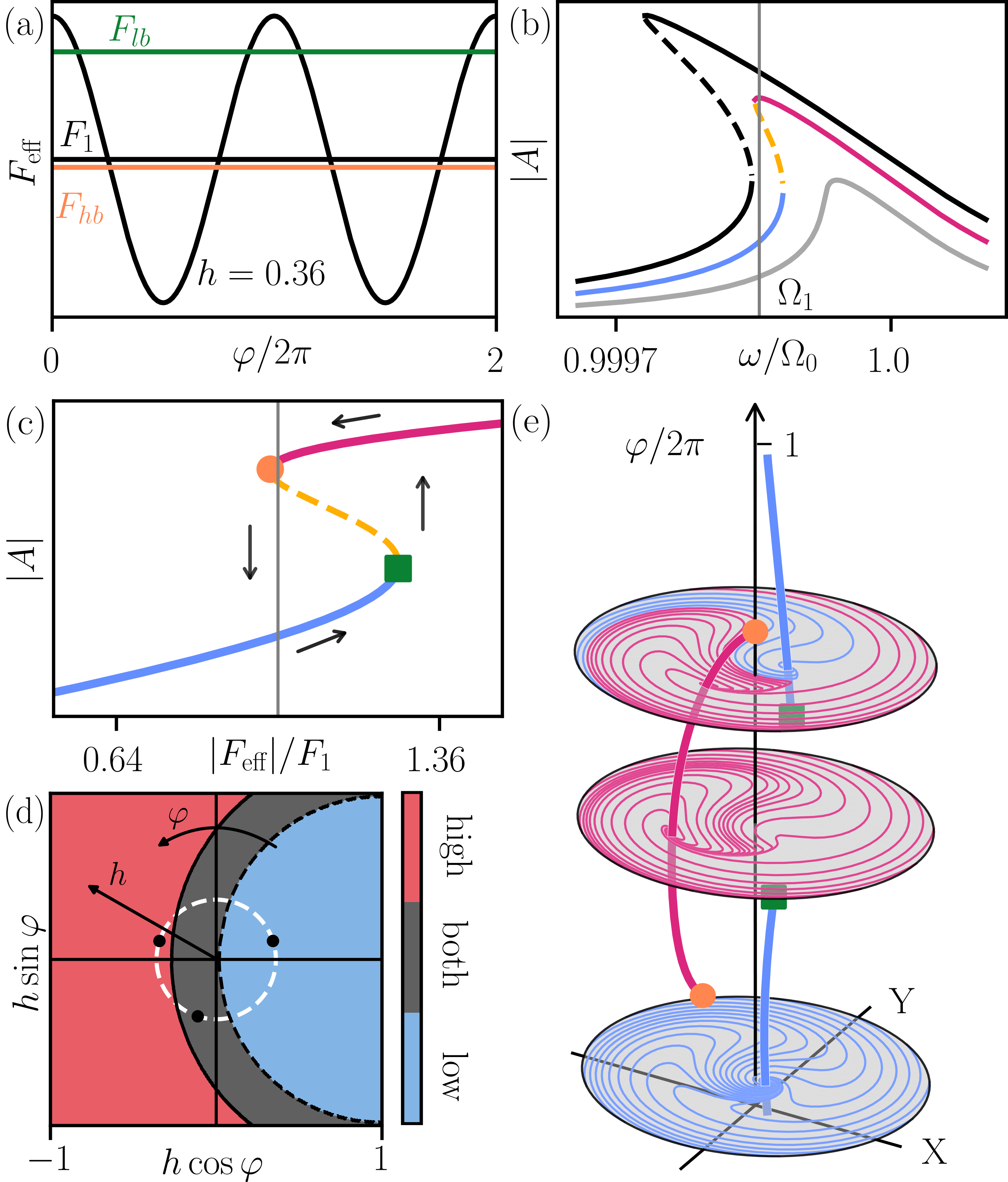}

\caption{The slow modulation picture. (a) The modulated effective drive amplitude, $F_{\mathrm{eff}}$, over two cycles of the relative phase $\varphi$ for $h=0.36 > h_{\mathrm{lb}}$. The horizontal lines indicate the bifurcation thresholds, $F_{\mathrm{lb}}$ (green) and $F_{\mathrm{hb}}$ (orange) [cf.~Eq.~\eqref{eq:duffing_bif}], and the primary tone's amplitude $F_{1}$ (black). (b) The stationary amplitude response as a function of drive frequency. The black (grey) curve shows the response at the maximum (minimum), $F_{\mathrm{eff}}^{\mathrm{max (min)}}$ for the modulation shown in (a). The coloured lines show the response to a single drive, i.e for $h=0$ as in Fig.~\ref{fig:fig1}(c). The vertical grey line marks the frequency of the primary tone, $\Omega_1$. (c) Bifurcation diagram against effective drive strength. As $|F_{\mathrm{eff}}|$ oscillates, the system follows the stable branches; arrows indicate the forced jumps between the low (blue) and high (pink) branches when thresholds are crossed. The orange circle (green square) marks $F_{hb\,(lb)}$. The vertical grey line marks $|F_{\mathrm{eff}}|=F_{1}$. (d) Quasistatic phase diagram [cf.~Eq.~\eqref{eq:2toneSlowFlowEqDuffing}] as a function of the relative drive strength $h$ and phase $\varphi(T)$. The grey region marks bistability, separated by bifurcation thresholds (solid and dashed black arcs) to  high- and low-branch only regions. Dashed white circle: trajectory for $h=0.36$. (e) Snapshot of vector flows at the three (black) points along the white trajectory in (d): monostable low (bottom), bistable (middle), and monostable high (top). The solid vertical arcs show the evolution of the blue (pink) low (high) branch as $\varphi$ varies. Parameters: $\Omega_{0}=2\pi$, $\Gamma=6.2\times10^{-4}$, and $F_{1}=4.81\times10^{-4}$.}
    \label{fig:Fmod_SharkShift}
\end{figure}

    To analyze the two-tone driven nonlinear resonator, we first adapt the methodology previously applied to the linear case under the assumption that the drive $F_1$ at frequency $\Omega_1$ generates the dominant response. This involves three key steps: (i) we employ the single-tone rotating ansatz, [cf.~Eq.~\eqref{eq:rotating_quadratures}];
    (ii) We write the decomposed drive [Eq.~\eqref{eq:ForceAfterFreqDecomposition}] as $F(t)=\mathrm{Re}\left[e^{i \left(\Omega_{1}t+\theta_{1}\right)}\left(1 + he^{i\left(\varphi(t)+\theta_{2}-\theta_{1}\right)}\right)\right] F_1$. Then, we assume that the time-evolving beating phase, $\varphi(t)=\Delta_{21} t$, varies slowly relative to the primary tone's period $T=2\pi/\Omega_{1}$, i.e., $\Delta_{21} \ll \Omega_{1}$; 
   (iii) We average over the timescale $T$, corresponding to the primary tone's period. Crucially, we still account for the slow time variation of the beating phase as $\varphi(T)$, and write the following equations of motion for the quadratures
    \begin{align}\label{eq:2toneSlowFlowEqDuffing}
    \begin{bmatrix}
    \dot{X} \\
    \dot{Y}
    \end{bmatrix}
    =
    \begin{bmatrix}
    -\frac{\Gamma}{2} & \frac{3 \alpha {A}^2 + 4 \tilde{\Delta}_{01}^2}{8 \Omega_1} \\
    -\frac{3 \alpha {A}^2 + 4 \tilde{\Delta}_{01}^2}{8 \Omega_1} & -\frac{\Gamma}{2}
    \end{bmatrix}
    \begin{bmatrix}
    X \\
    Y
    \end{bmatrix}
    +
    \vec{F}_{\mathrm{eff}} \,,
    \end{align}
     where 
    \begin{equation}\label{eq:mod_effective_drive}
    \vec{F}_{\mathrm{eff}}=\frac{F_1}{2\Omega_1}\begin{bmatrix}
     \sin \theta_1 + h \sin(\varphi(T) + \theta_2) \\
    \cos \theta_1 + h \cos(\varphi(T) + \theta_2)
    \end{bmatrix}\,.
    \end{equation}
    Importantly, this approach differs from the single-tone HBM used in Sec.~\ref{sec:OneNLresp}, where a single-harmonic ansatz would average out responses at frequencies other than $\Omega_1$ and thus fail to capture the dynamics induced by the secondary tone through $\varphi(T)=\Delta_{21} T$.

    The time-dependent phase $\varphi(T)$ in $\vec{F}_{\mathrm{eff}}$ [cf.~Eq.~\eqref{eq:mod_effective_drive}] implies that the secondary tone acts as an amplitude modulation of the primary tone in the $\Omega_{1}$-rotating frame, see Fig.~\ref{fig:Fmod_SharkShift}(a). This modulation's effect is most clearly understood in the slow limit ($\Delta_{21}\rightarrow0$), where $\varphi$ becomes a quasistatic parameter that alters the system's stability landscape through the modulated drive amplitude,    
    \begin{equation}\label{eq:effective_drive_length}
    |F_{\mathrm{eff}}|=F_{1}\sqrt{1+h^2+2h\cos{(\theta_{2}-\theta_{1}-\varphi(T))}}\,,
  \end{equation}
    which oscillates between $F_{\mathrm{eff}, -}\equiv F_1(1-h)$ and $F_{\mathrm{eff}, +}\equiv F_1(1 + h)$ as $\varphi$ evolves. This modulation changes the rotating potential landscape, and hence the response of the system, see Fig.~\ref{fig:Fmod_SharkShift}(b). 
    
    Hereafter, we assume the system is initialized in the low-amplitude stationary state of the single-tone problem (i.e., when $F_{2}=0$). The modulated amplitude of $F_{\mathrm{eff}}$ can trigger a jump if $|F_{\mathrm{eff}}|$ is large enough for the system to cross the lower branch bifurcation at $\Omega_1$, see Fig.~\ref{fig:Fmod_SharkShift}(c). Thus, for the jumps to occur as $\varphi$ is varied, $F_{\mathrm{eff}, +}$ must satisfy or overshoot the discriminant condition for bifurcation, [cf.~Eq.~\eqref{eq:duffing_bif}]. This leads to a critical value $h_{\mathrm{lb}}$ for the lower branch when $F_{\mathrm{eff}, +}=F_{\mathrm{lb}}\equiv F_{1}(1+h_{\mathrm{lb}})$ exactly satisfies Eq.~\eqref{eq:duffing_bif}. For $h>h_{\mathrm{lb}}$, the system jumps when the low-amplitude attractor loses stability as $\varphi$ is varied and moves to the high-amplitude attractor, see Fig.~\ref{fig:Fmod_SharkShift}(c). Similarly, the high amplitude branch bifurcates when $F_{\mathrm{eff}, -}\leq F_{\mathrm{hb}}\equiv F_{1}(1-h_{\mathrm{hb}})$ and the system moves back to the low-amplitude attractor. 
    
    These \textit{inter-attractor} jumps can be tracked by looking at the change in the stability of the attractors for varying $\varphi$ and $h$ in Fig.~\ref{fig:Fmod_SharkShift}(d): for a circular trajectory at fixed $h>h_{\mathrm{lb}}$, as $\varphi$ varies and $F_{\mathrm{eff}}$ changes, the system crosses the bifurcation point of the lower branch and consequently jumps to the available stationary branch and moves across stability regions where the low, high or both branches are stable as in Fig.~\ref{fig:Fmod_SharkShift}(e). This agrees with the recent experiment and analysis presented in Ref.~\cite{catalini_slow_2025}. However, when the beating between the tones becomes faster ($|\Delta_{21}|\geq\Gamma$), the quasistatic picture breaks down, necessitating a more comprehensive analysis to capture the system's dynamics. 

    \begin{figure}[t]
        \centering
        \includegraphics[width=\columnwidth]{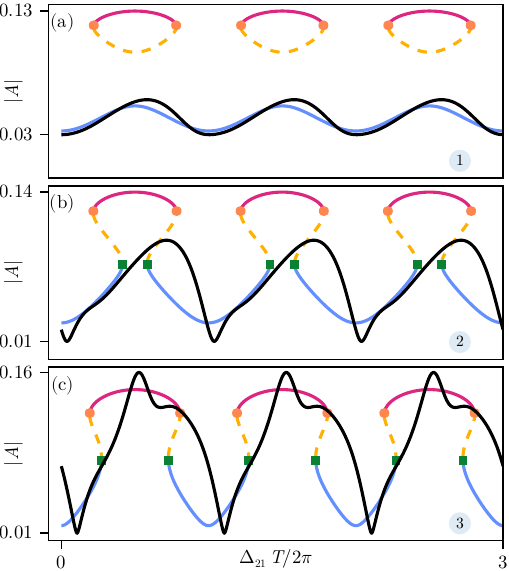}
        \caption{The time-domain response $A(T)$ (black curve), simulated over three modulation cycles, attempting to follow the stationary states calculated from the quasistatic Eq.~\eqref{eq:2toneSlowFlowEqDuffing} - pink (blue) solid for high (low) amplitude state, yellow dashed line for the saddle, with $\Delta_{21}=0.42\Gamma$ for (a) $h=0.1$, (b) $h=0.33$ and (c) $h=0.65$. }
        \label{fig:Fixed_Detuning_Trajectories}
    \end{figure}

    Building on Ref.~\cite{catalini_slow_2025}, the decisive timescale for applying the quasi-static picture is the resonator ringdown, $\tau = 1/\Gamma$, which low-pass filters the beat-note modulation at $ |\Delta_{21}| $. When $|\Delta_{21}| \ll \Gamma$ (beat period $\gg \tau$), the modulation is quasistatic: the vector-flow evolves slowly and the state continuously follows the stationary branches until the branch bifurcates [Fig.~\ref{fig:Fmod_SharkShift}(b)]. This enforces completed transitions between the low- and high-amplitude Duffing attractors once per cycle. Conversely, when $|\Delta_{21}| \lesssim \Gamma$ (beat period $\lesssim \tau$), the modulation is fast: dissipation cannot relax the response to the changing attractor before it merges with the saddle, so the trajectory lags the changing flow and remains trapped near the initial basin, producing small loops without completed inter-attractor jumps despite an unchanged instantaneous vector-flow with respect to the quasistatic case. Thus, naively, $\tau$ seems to set the boundary between slow transitions in Fig.~\ref{fig:Fmod_SharkShift}(c) and fast, lagging dynamics. In the following, we show that this initial intuition from Ref.~\cite{catalini_slow_2025} does not depict the full picture.
    
\subsection{\label{sec:Dynamical phase transitions and phase trapping}Dynamical phase transitions}

\begin{figure}[ht]
    \centering
    \includegraphics[width=\columnwidth]{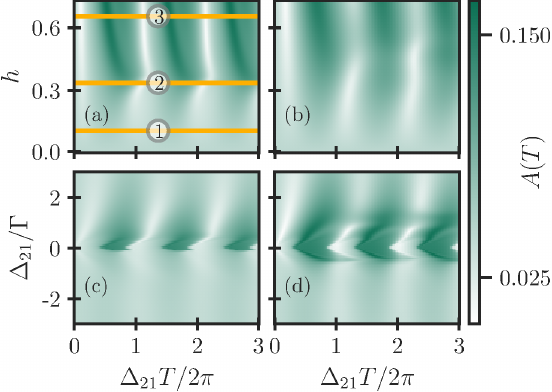}

    \caption{Time-domain response amplitude $A(T)$ over three modulation cycles. (a) Slow modulation regime ($\Delta_{21}=0.42\Gamma$): Numbered lines indicate cuts for $h=0.1$ (stable low), $h=0.33$ (deviation), and $h=0.65$ (switching), corresponding to the trajectories in Fig.~\ref{fig:Fixed_Detuning_Trajectories}. (b) Same as (a) with fixed high detuning $\Delta_{21}=0.91\Gamma$. The response exhibits significant phase lag and smooth changes compared with (a). (c) Fixed $h=0.36$ with varying detuning: $\Delta_{21}>0$ triggers larger amplitudes in response to the modulation than $\Delta_{21}<0$. (d) Same as (c) with $h=0.71$. Sharp dark green regions signal large amplitudes corresponding to jumps to the high amplitude state.}
    \label{fig:fig2_AmpTimeColormaps}
\end{figure}

    How do the trajectories change when the modulation is not infinitely slow? To answer this, we numerically solve Eq.~\eqref{eq:2toneSlowFlowEqDuffing}, initializing the system in the low-amplitude stationary state of the single-tone case ($h=0$) before introducing the secondary tone at $T=0$. We look at the dynamics at a fixed, finite detuning $\Delta_{21}$ to understand the effect of the relative strength $h$ on these transitions. The resulting amplitude dynamics $A(T)$ are contingent on $h$, see Fig.~\ref{fig:Fixed_Detuning_Trajectories}. For $h<h_{\mathrm{lb}}$, the initial lower amplitude state is stable throughout the modulation and the system stays close to it, see Fig.~\ref{fig:Fixed_Detuning_Trajectories}(a). However, for $h>h_{\mathrm{lb}}$, we see two distinct types of behaviours. In Fig.~\ref{fig:Fixed_Detuning_Trajectories}(b), the amplitude of the system rises above the low amplitude branch after it bifurcates, but never reaches the high-amplitude branch. In contrast, in Fig.~\ref{fig:Fixed_Detuning_Trajectories}(c), the system rises above the low-amplitude branch and reaches the high-amplitude branch during the modulation cycle.
    
    To look at the complex dependence of the dynamics on $\Delta_{21}$ and $h$, we fix one parameter at small and large values and sweep across the other, see Fig.~\ref{fig:fig2_AmpTimeColormaps}. The sharp jumps in $A(T)$ for $h>h_{\mathrm{lb}}$ in the slow beating limit have the same period as the drive modulation, see Fig.~\ref{fig:fig2_AmpTimeColormaps}(a). However, as the beating becomes faster ($\Delta_{21} \sim \Gamma$) the response accumulates a significant phase lag and changes smoothly, see Fig.~\ref{fig:fig2_AmpTimeColormaps}(b). Furthermore, these dynamics are asymmetric with respect to $\Delta_{21}$: positively detuned secondary tones ($\Delta_{21}>0$) trigger large amplitude variations over a wider range of detunings and lead to higher peak values than their negatively detuned ($\Delta_{21}<0$) counterparts, see Fig.~\ref{fig:fig2_AmpTimeColormaps}(c) and (d). The diverse nature of these amplitude dynamics across parameter space motivates using an effective quantity to distinguish the distinct dynamical regimes and track the transitions across these regimes.

    \begin{figure*}[ht!]
    \centering
    \includegraphics[]{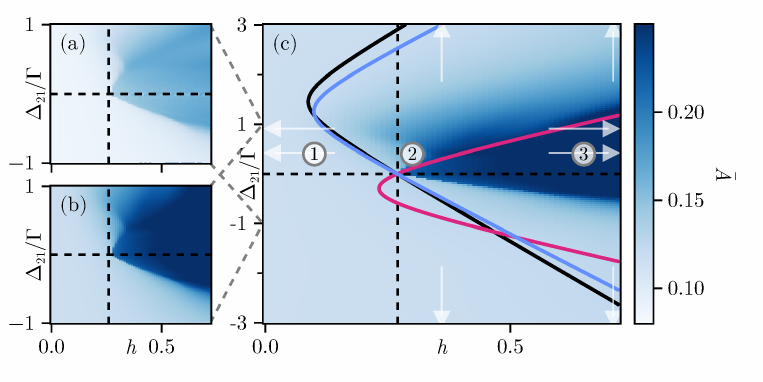}

    \caption{The average amplitude in one modulation cycle, $\bar{A}$, as a function of relative strength $h$ and frequency $\Delta_{21}$.~(a) Experimental data (adapted from Ref.~\cite{catalini_slow_2025}, cf.~Appendix~\ref{sec:experimental_setup}). (b) Numerical simulation of Eq.~\eqref{eq:2toneSlowFlowEqDuffing}. (c) Extended simulation range with analytical transition boundaries overlaid. The \textbf{black line} represents the instability threshold of the low-amplitude branch based on the bare resonance [cf.~Eq.~\eqref{eq:effective_Linear_drives_Omega1}]. The \textbf{pink line} represents the transition threshold incorporating the resonance frequency shift due to the high amplitude state [cf.~Eq.~\eqref{eq:renormalised_filter_effective_high}]. The \textbf{blue line} uses the renormalized frequency of the low-amplitude state [cf.~Eq.~\eqref{eq:renormalised_filter_effective_low}]. Markers \protect\circled{1}-\protect\circled{3} correspond to the regimes illustrated in Fig.~\ref{fig:Fixed_Detuning_Trajectories}. The horizontal arrows correspond to the fixed $\Delta_{21}$ cuts and the vertical arrows correspond to fixed $h$ cuts in Fig.~\ref{fig:fig2_AmpTimeColormaps}.}
    \label{fig:fig4_AvgAmp_PhaseDiagram}
    \end{figure*}

    We introduce the time-averaged amplitude over one drive modulation cycle, $\bar{A}$, defined as:
    \begin{equation}\label{eq:avg_amplitude_def}
    \bar{A}(h, \Delta_{21}) = \frac{|\Delta_{21}|}{2\pi} \int_{T_0}^{T_0 + 2\pi/|\Delta_{21}|}  \,A(T; h, \Delta_{21})\,\rm{d}T ,
    \end{equation}
    with amplitude $A(T)$ obtained from the time evolution of Eq.~\eqref{eq:2toneSlowFlowEqDuffing}. The cycle-averaged amplitude $\bar{A}$ serves as an order parameter for characterizing the system's long-term behaviour. In Fig.~\ref{fig:fig4_AvgAmp_PhaseDiagram}, we present the phase diagram of $\bar{A}$ as a function of $h$ and $\Delta_{21}$, obtained from (a) experimental data measured on the setup [cf.~Ref.~\cite{catalini_slow_2025} and Appendix.~\ref{sec:experimental_setup}] and (b) numerical simulations. The $\bar{A}$ values of (a) are slightly lower than those of (b), suggesting that higher order nonlinearities may result in a saturated experimental response. We observe excellent agreement between (a) and (b) in terms of the overall shape of the high-$\bar{A}$ region, which validates our theoretical model, Eq~\eqref{eq:2toneSlowFlowEqDuffing}. 
    
    In Fig.~\ref{fig:fig4_AvgAmp_PhaseDiagram}, we clearly distinguish between two dynamical regimes: dynamics confined to the low-amplitude attractor (low $\bar{A}$, light blue regions), and dynamics involving the high-amplitude attractor (high $\bar{A}$, dark regions), which encompass inter-attractor bursting oscillations and trajectories localized around the high-amplitude attractor. Interestingly, the transition boundary separating these two regimes is asymmetric: it is approximately linear in $h$ for $\Delta_{21} < 0$ and appears parabolic for $\Delta_{21} > 0$. This representation clearly demarcates parameter regions associated with qualitatively different behaviours, particularly identifying the onset of inter-attractor jumps. Analytically determining this boundary in a general closed form, analogous to the single-tone approach [Eq.~\eqref{eq:AmplitudePolynomial}, Eq.~\eqref{eq:duffing_bif}], is unfeasible due to the non-autonomous (i.e., explicitly time-dependent) nature of the two-tone system [Eq.~\eqref{eq:2toneSlowFlowEqDuffing}]. The critical question then becomes: can we approximate the boundary for these dynamical phase transitions, specifically the $(h, \Delta_{21})$ conditions that precipitate such transitions?

    \section{\label{sec:Phase Transition Boundaries}Phase Transition Boundaries}

In Sec.~\ref{sec:TwoNLresp}, we observed that the secondary tone can lead to modulation and instabilities of the stationary response to the primary tone. In the slow modulation limit, we used Eq.~\eqref{eq:2toneSlowFlowEqDuffing} to argue that the response at frequency $\Omega_1$ experiences a slow modulation due to the time-dependent effective drive amplitude in Eq.~\eqref{eq:mod_effective_drive} 
and thus can exceed a threshold beyond which the initial single-tone branch destabilizes. However, this approach does not take into account the time it takes the system to respond to changes in the amplitude of the external drive, of order $\Gamma^{-1}$. In Fourier domain, this approach entails
that we neglected the fact that the resonator has a different susceptibility to the secondary tone compared to the primary tone, cf.~Fig.~\ref{fig:fig1}(b). As such, we cannot account for why the threshold in Fig.~\ref{fig:fig4_AvgAmp_PhaseDiagram} increases with $|\Delta_{21}|$, nor why this increase occurs in an asymmetric fashion with respect to $\Delta_{21}$.

\begin{table*}[t] 
\centering \small \setlength{\tabcolsep}{5pt} \renewcommand{\arraystretch}{1.5} %
\begin{tabular}
{@{}>{\centering\arraybackslash}p{0.05\linewidth} @{}>{\centering\arraybackslash}p{0.15\linewidth} @{}>{\centering\arraybackslash}p{0.5\linewidth} @{}>{\centering\arraybackslash}p{0.2\linewidth}@{}} \toprule \textbf{No.} & \textbf{Start branch} & \textbf{Regime} & \textbf{Outcome} \\ \midrule 
\circled{1} & lb & ${F_\mathrm{eff}}_{\raisebox{-0.3ex}{$\scriptscriptstyle \Omega_{\mathrm{l}}, +$}} < F_{\mathrm{lb}}$ & stays lb \\ 
\circled{2} & lb & ${F_\mathrm{eff}}_{\raisebox{-0.3ex}{$\scriptscriptstyle \Omega_{\mathrm{l}}, +$}} < F_{\mathrm{lb}} < {F_\mathrm{eff}}_{\raisebox{-0.3ex}{$\scriptscriptstyle \Omega_{\mathrm{h}}, +$}}$ & lb + deviation \\ 
\circled{3} & lb & $F_{\mathrm{lb}} < {F_\mathrm{eff}}_{\raisebox{-0.3ex}{$\scriptscriptstyle \Omega_{\mathrm{l}}, +$}}, {F_\mathrm{eff}}_{\raisebox{-0.3ex}{$\scriptscriptstyle \Omega_{\mathrm{h}}, +$}} \text{ and } {F_\mathrm{eff}}_{\raisebox{-0.3ex}{$\scriptscriptstyle \Omega_{\mathrm{h}}, -$}} < F_{\mathrm{hb}}$ & lb $\leftrightarrow$ hb \\ 
\circled{4} & lb & $F_{\mathrm{lb}} < {F_\mathrm{eff}}_{\raisebox{-0.3ex}{$\scriptscriptstyle \Omega_{\mathrm{l}}, +$}}, {F_\mathrm{eff}}_{\raisebox{-0.3ex}{$\scriptscriptstyle \Omega_{\mathrm{h}}, +$}} \text{ and } {F_\mathrm{eff}}_{\raisebox{-0.3ex}{$\scriptscriptstyle \Omega_{\mathrm{h}}, -$}} > F_{\mathrm{hb}}$ & moves to hb and stays \\ \midrule %
\circled{5} & hb & $F_{\mathrm{hb}} < {F_\mathrm{eff}}_{\raisebox{-0.3ex}{$\scriptscriptstyle \Omega_{\mathrm{l}}, -$}}, {F_\mathrm{eff}}_{\raisebox{-0.3ex}{$\scriptscriptstyle \Omega_{\mathrm{h}}, -$}}$ & stays hb \\ 
\circled{6} & hb & ${F_\mathrm{eff}}_{\raisebox{-0.3ex}{$\scriptscriptstyle \Omega_{\mathrm{h}}, -$}} < F_{\mathrm{hb}} < {F_\mathrm{eff}}_{\raisebox{-0.3ex}{$\scriptscriptstyle \Omega_{\mathrm{l}}, -$}}$ & hb + deviation \\ 
\circled{7} & hb & ${F_\mathrm{eff}}_{\raisebox{-0.3ex}{$\scriptscriptstyle \Omega_{\mathrm{l}}, -$}}, {F_\mathrm{eff}}_{\raisebox{-0.3ex}{$\scriptscriptstyle \Omega_{\mathrm{h}}, -$}} < F_{\mathrm{hb}} \text{ and } {F_\mathrm{eff}}_{\raisebox{-0.3ex}{$\scriptscriptstyle \Omega_{\mathrm{l}}, +$}}, {F_\mathrm{eff}}_{\raisebox{-0.3ex}{$\scriptscriptstyle \Omega_{\mathrm{h}}, +$}} > F_{\mathrm{lb}}$ & hb $\leftrightarrow$ lb \\ 
\circled{8} & hb & ${F_\mathrm{eff}}_{\raisebox{-0.3ex}{$\scriptscriptstyle \Omega_{\mathrm{l}}, -$}}, {F_\mathrm{eff}}_{\raisebox{-0.3ex}{$\scriptscriptstyle \Omega_{\mathrm{h}}, -$}} < F_{\mathrm{hb}} \text{ and } {F_\mathrm{eff}}_{\raisebox{-0.3ex}{$\scriptscriptstyle \Omega_{\mathrm{l}}, +$}}, {F_\mathrm{eff}}_{\raisebox{-0.3ex}{$\scriptscriptstyle \Omega_{\mathrm{h}}, +$}} < F_{\mathrm{lb}}$ & moves to lb and stays \\ 
\bottomrule 
\end{tabular}

\caption{\label{tab:transition_regimes}This table summarizes the different dynamical regions assuming a fixed primary tone in the bistability region, \protect\circled{1} - \protect\circled{8}, differentiated by the starting branch [low(high) branch as lb(hb)] and the detuning and strength of the secondary tone. The properties of the secondary tone lead to different effective filtered drive strengths during modulation, see Eqs.~\eqref{eq:effective_Linear_drives_Omega1} and \eqref{eq:renormalised_filter_effective_high}, which we call regimes. $F_{\mathrm{hb}}$, $F_{\mathrm{lb}}$ are the bifurcation thresholds for the high and lower branch obtained in Eq.~\eqref{eq:duffing_bif}. ${F_\mathrm{eff}}_{\raisebox{-0.3ex}{$\scriptscriptstyle \Omega_{\mathrm{h}}, \pm$}}$ is the maximum (minimum) renormalized filtered effective drive obtained in Eq.~\eqref{eq:renormalised_filter_effective_high} that takes into account the dispersive shift [cf.~Eq.~\eqref{eq:disp_shift}] due to the high amplitude state, $A_{\text{high}}$. ${F_\mathrm{eff}}_{\raisebox{-0.3ex}{$\scriptscriptstyle \Omega_{\mathrm{l}}, \pm$}}$ is the renormalized filtered effective drive that incorporates resonance renormalization due to the low amplitude state, $A_{\text{low}}$. \protect\circled{1} - \protect\circled{3} refer to the cases presented in Figs.~\ref{fig:Fixed_Detuning_Trajectories}(a)-(c) and marked on the phase diagram in Fig.~\ref{fig:fig4_AvgAmp_PhaseDiagram}(c).Cases \protect\circled{5}-\protect\circled{7} are presented in Figs.~\ref{fig:nearlb_fixed_trajs}(a)-(c) and are marked on the phase diagram in Fig.~\ref{fig:nearlb_starthi}.}
\end{table*}

To model the response of the system to the amplitude modulation of the external drive, we turn back to the power of the linear system at $\Omega_{1}$, modulated due to the secondary tone, cf.~Eq.~\eqref{eq:linearPoweratPumptimedependence} in Sec.~\ref{sec:TwoLinResp}. Indeed, the instability of the lower branch depends more on the amount of power actually received by the system, rather than the modulation amplitude of the external drive. Received power and drive amplitude are often closely associated, but under the conditions that we consider, they can diverge strongly: consider, as an example, a high-amplitude modulation which is very fast, such that the system is unable to respond in time. The average power received by the system can then be very low, in spite of the high peak drive amplitude, cf.~Fig.~\ref{fig:fig2_AmpTimeColormaps}(d) for $\Delta_{21}/\Gamma = 2$.

In this Section, we incorporate the information from linear response, i.e., the susceptibility function, on top of the stationary nonlinear motion of the Duffing resonator, to approximate the mechanism that allows for inter-attractor trajectories. We first propose that the two-tone linear response power absorbed by the resonator, see Eq.~\eqref{eq:linearPoweratPumptimedependence} serves as a good first approximation to the two-tone driven Duffing resonator. Secondly, we use this approximation to evaluate the filtered effective drive experienced at $\Omega_{1}$. To this end, we use Eq.~\eqref{eq:linearPoweratPumptimedependence} to calculate the maximum and minimum linear response power, $A^2_{\raisebox{-0.3ex}{$\scriptscriptstyle \Omega_0, +$}}$ and $A^2_{\raisebox{-0.3ex}{$\scriptscriptstyle \Omega_0, -$}}$ respectively, achieved in a modulation cycle. Using the identity $\max_T[a_{\raisebox{-0.3ex}{$\scriptscriptstyle \Omega_0$}}\cos(\Delta_{21} T)+b_{\raisebox{-0.3ex}{$\scriptscriptstyle \Omega_0$}}\sin(\Delta_{21} T)]=\sqrt{a_{\raisebox{-0.3ex}{$\scriptscriptstyle \Omega_0$}}^2+b_{\raisebox{-0.3ex}{$\scriptscriptstyle \Omega_0$}}^2}$ at $T=\arctan(b_{\raisebox{-0.3ex}{$\scriptscriptstyle \Omega_0$}}/a_{\raisebox{-0.3ex}{$\scriptscriptstyle \Omega_0$}})/\Delta_{21}$, the expressions for the extrema read
    \begin{multline}\label{eq:LinearPowerMaxMin}
        A_{\raisebox{-0.3ex}{$\scriptscriptstyle \Omega_0,\pm$}}^2 = F_1^2(|\chi_{1}|^2\!+\!h^2|\chi_{2}|^2\!)\\\pm\!2F_{1}^2h|\chi_{1}|^2|\chi_{2}|^2\!\sqrt{a_{\raisebox{-0.3ex}{$\scriptscriptstyle \Omega_0$}}^2+b_{\raisebox{-0.3ex}{$\scriptscriptstyle \Omega_0$}}^2},
    \end{multline}
    where $a_{\raisebox{-0.3ex}{$\scriptscriptstyle \Omega_0$}}$, $b_{\raisebox{-0.3ex}{$\scriptscriptstyle \Omega_0$}}$ are given by Eqs.~\eqref{eq:shorthand_a} and~\eqref{eq:shorthand_b}.
    This maximum and minimum linear response power crucially depends on the strength and detuning of the secondary tone. Note that Eq.~\eqref{eq:LinearPowerMaxMin} describes the extremal received power at $\Omega_{1}$ due to an off-resonant or detuned secondary tone. In contrast to Eq.~\eqref{eq:2toneSlowFlowEqDuffing}, it ignores the Duffing nonlinearity but incorporates the effect of detuning. 
    
    Starting from Eq.~\eqref{eq:LinearPowerMaxMin}, we use the linear susceptibility, $\chi_{1}$ [cf.~Eq.~\eqref{eq:SHOResponseLab} for a single tone drive], to formulate an in-phase effective (filtered) force ${F_\mathrm{eff}}_{\raisebox{-0.3ex}{$\scriptscriptstyle \Omega_0, \pm$}}$ that produces the same stationary response $A^2_{\raisebox{-0.3ex}{$\scriptscriptstyle \Omega_0,\pm$}}$, at $\Omega_{1}$,
    \begin{equation}\label{eq:effective_Linear_drives_Omega1}
    	{F^2_{\mathrm{eff}}}_{\raisebox{-0.3ex}{$\scriptscriptstyle \Omega_0, \pm$}} = A^2_{\raisebox{-0.3ex}{$\scriptscriptstyle \Omega_0,\pm$}}/|\chi_{1}|^2\,.
    \end{equation}
    This, in turn, enables us to find the effective drive amplitude required for a bifurcation at $\Omega_{1}$, bringing us back to a similar threshold analysis as we explored in Sec.~\ref{sec:TwoNLresp}.

    The comparison between the effective drive strengths obtained in Eq.~\eqref{eq:effective_Linear_drives_Omega1} and the critical drive amplitudes for bifurcation obtained in Eq.~\eqref{eq:duffing_bif} provides us with the required force for jumping away from the lower branch towards the higher branch, i.e., ${F_{\mathrm{eff}}}_{\raisebox{-0.3ex}{$\scriptscriptstyle \Omega_0, +$}} \geq F_{\mathrm{lb}}$, cf.~Fig.~\ref{fig:Fmod_SharkShift}(c). The threshold condition ${F_{\mathrm{eff}}}_{\raisebox{-0.3ex}{$\scriptscriptstyle \Omega_0, +$}}=F_{\mathrm{lb}}$ leads to the black transition threshold in Fig.~\ref{fig:fig4_AvgAmp_PhaseDiagram}(c).  
    In Sec.~\ref{sec:TwoNLresp}, the slow modulation allowed the system to reach the higher branch simply by leaving the lower one. With fast modulation, however, we must check that the system can still absorb sufficient power from the secondary tone during the jump. Indeed, as the system leaves the lower branch and approaches the higher branch, the oscillation amplitude grows, and the resonance frequency shifts with amplitude according to~\cite{eichler_classical_2023}
    \begin{equation}\label{eq:disp_shift}
    \Omega_{\mathrm{r}}(A) = \Omega_0 (1 + \frac{3\alpha}{8\Omega_0^2}A^2)\,,
    \end{equation}
    detuning the system from the secondary tone. This ``dynamical'' Duffing frequency shift~\eqref{eq:disp_shift} changes the filtered effective drive strength~\eqref{eq:effective_Linear_drives_Omega1} and potentially prevents the transition to the higher branch. Therefore, completing the transition requires enough power to leave the lower branch and overcome resonance renormalization while keeping the absorbed power, and thus the filtered force in Eq.~\eqref{eq:effective_Linear_drives_Omega1}, above the $F_{\mathrm{lb}}$ threshold. We incorporate a conservative estimate of this effect using the amplitude of the higher branch, $A_{\text{high}}$, which is the target state of the transition once the lower branch destabilizes, cf.~Fig.~\ref{fig:Fmod_SharkShift}(c), to calculate a renormalized filtered effective drive, 
    \begin{equation}\label{eq:renormalised_filter_effective_high}
        {F^2_{\mathrm{eff}}}_{\raisebox{-0.3ex}{$\scriptscriptstyle \Omega_{\mathrm{h}}, \pm$}} = A^2_{\raisebox{-0.3ex}{$\scriptscriptstyle \Omega_{\mathrm{h}}, \pm$}}/|\chi_{\raisebox{-0.3ex}{$\scriptscriptstyle \Omega_{\mathrm{h}}$}}(\Omega_1)|^2\,,
    \end{equation}
    by combining Eqs.~\eqref{eq:LinearPowerMaxMin} and~\eqref{eq:disp_shift}, defining the renormalized susceptibility $\chi_{\raisebox{-0.3ex}{$\scriptscriptstyle \Omega_{\mathrm{h}}$}}(\Omega_1)$ and the renormalized absorbed  power extrema $A^2_{\raisebox{-0.3ex}{$\scriptscriptstyle \Omega_{\mathrm{h}}, \pm$}}$ where $\Omega_{\mathrm{h}}\equiv\Omega_{\mathrm{\mathrm{r}}}(A_{\mathrm{high}})$. The condition ${F_{\mathrm{eff}}}_{\raisebox{-0.3ex}{$\scriptscriptstyle \Omega_{\mathrm{h}}, +$}}=F_{\mathrm{lb}}$ yields the transition boundary for the system to leave the lower branch and reach the higher branch during the modulation cycle, see pink curve in Fig.~\ref{fig:fig4_AvgAmp_PhaseDiagram}(c).

In summary, two distinct thresholds govern the transition dynamics: the black line in Fig.~\ref{fig:fig4_AvgAmp_PhaseDiagram}(c) identifies the condition where the lower branch becomes unstable and the system is driven away from it, while the pink line identifies the regime where absorbed power from the secondary tone remains sufficient to cross the lower-branch instability and reach the higher-branch, despite  the amplitude-dependent frequency renormalization [cf.~Eqs.~\eqref{eq:disp_shift} and~\eqref{eq:renormalised_filter_effective_high}]. The interplay between these two thresholds explains the different dynamical regimes, which we illustrate with three representative trajectories: \circled{1} lies below both thresholds, where the trajectory remains confined to the lower branch as it remains stable throughout the modulation cycle, see Fig.~\ref{fig:Fixed_Detuning_Trajectories}(a); \circled{2} lies above the black but below the pink threshold, where the system departs from the lower branch but fails to reach the higher branch due to resonance frequency renormalization, see Fig.~\ref{fig:Fixed_Detuning_Trajectories}(b); and \circled{3} lies above both thresholds, where the system possesses enough power to both destabilize the lower branch and complete the jump, successfully reaching the higher branch during its cycle, see Fig.~\ref{fig:Fixed_Detuning_Trajectories}(c). Importantly, the pink line provides a good approximation to the parabolic instability boundary observed in Fig.~\ref{fig:fig4_AvgAmp_PhaseDiagram}(c) for positive detuning ($\Delta_{21} > 0$), marking the success of our analytical approach.

Although our threshold estimates capture the main features of the two-tone dynamics, notable deviations arise for negative detuning ($\Delta_{21} < 0$), where the black transition threshold deviates significantly from the experimental/simulated transition. To improve this, we include resonance renormalization arising from the finite amplitude of the lower branch, $A_{\text{low}}$ close to the system's initial state. We use Eq.~\eqref{eq:disp_shift} to define $\Omega_{\mathrm{l}}\equiv \Omega_{\mathrm{r}}(A_{\mathrm{low}})$ and correct Eq.~\eqref{eq:effective_Linear_drives_Omega1} to get 
\begin{equation}\label{eq:renormalised_filter_effective_low}
    {F_\mathrm{eff}}_{\raisebox{-0.3ex}{$\scriptscriptstyle \Omega_{\mathrm{l}}, \pm$}} = A^2_{\raisebox{-0.3ex}{$\scriptscriptstyle \Omega_{\mathrm{l}}, \pm$}}/|\chi_{\raisebox{-0.3ex}{$\scriptscriptstyle \Omega_{\mathrm{l}}$}}(\Omega_1)|^2\,,
\end{equation}
the renormalized filtered effective drive incorporating the shift due to $A_{\mathrm{low}}$.
The resulting blue threshold line in Fig.~\ref{fig:fig4_AvgAmp_PhaseDiagram}(c) is closer to the observed transition for negative detunings ($\Delta_{21} < 0$). This points to the relevance of higher-order renormalization schemes, which could eventually reconcile the effective-drive model with the full nonlinear dynamics. Remaining discrepancies can be traced to (i) the coarse-grained nature of the time-averaged order parameter~\eqref{eq:avg_amplitude_def}, which tends to smear transitions, and (ii) the simplifying assumptions underlying our analytical treatment, which are intended to capture the qualitative threshold rather than its exact position. The latter motivates the study of the problem using a multifrequency ansatz.

Beyond the specific case studied here, other combinations of thresholds and trajectories are possible depending on the initial starting stationary state and the chosen primary tone detuning. For instance, starting from the higher branch, the system will leave the higher branch towards the lower branch if ${F_\mathrm{eff}}_{\raisebox{-0.3ex}{$\scriptscriptstyle \Omega_{\mathrm{h}}, -$}}\leq F_{\mathrm{hb}}$ and will reach the lower branch if ${F_{\mathrm{eff}}}_{\raisebox{-0.3ex}{$\scriptscriptstyle \Omega_0, -$}}\leq F_{\mathrm{hb}}$. These additional scenarios are systematically summarized in Table~\ref{tab:transition_regimes}, which maps the inequalities between ${F_\mathrm{eff}}_{\raisebox{-0.3ex}{$\scriptscriptstyle \Omega_{\mathrm{h}}, +$}}$, ${F_\mathrm{eff}}_{\raisebox{-0.3ex}{$\scriptscriptstyle \Omega_{\mathrm{h}}, -$}}$, $F_{\mathrm{lb}}$, and $F_{\mathrm{hb}}$ onto the corresponding dynamical outcomes. This overview highlights that the observed behaviour is just one instance within a broader set of possible instability regimes, underscoring the richness of the system’s driven dynamics, cf.~Appendix~\ref{sec:additional_cases} for additional examples. There, too, depending on the particular parameters used, the transition thresholds obtained from our model can deviate from the boundary observed in simulations. These deviations also trace back to the simplifying assumption of using a two-tone driven linear response to approximate the two-tone driven Duffing resonator's nonlinear response.

\section{\label{sec:Extension of Ansatz}Extension of the Ansatz}

\begin{figure}[t]
    \centering
    \includegraphics[]{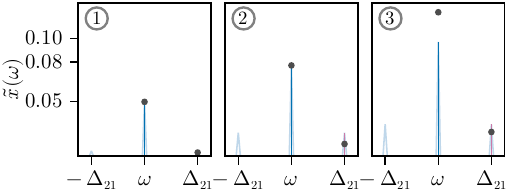}
    \caption{Fourier spectra of the trajectories from Fig.~\ref{fig:Fixed_Detuning_Trajectories}. The peaks at $\omega=0$ (blue) and $\omega=\Delta_{21}$ (pink) correspond to the response components at the primary tone $\Omega_1$ and secondary tone $\Omega_2$, respectively, in the $\Omega_1$-rotating frame. Grey markers indicate the stationary state amplitudes $A_1$ and $A_2$ predicted by the multifrequency HBM ansatz [Eqs.~\eqref{eq:extended_two_tone_HarmonicEqns}]. The other parameters used are $\Omega_{0}=2\pi$, $\Gamma=6.2\times10^{-4}$, $F_{1}=4.81\times10^{-4}$, and $\Delta_{10}=-9.02\times10^{-4}$.}
    \label{fig:fft_trajectories}
\end{figure}

In Sections~\ref{sec:TwoNLresp} and~\ref{sec:Phase Transition Boundaries}, we analyzed how the secondary tone modifies the stationary response of the primary tone. The analysis stemmed from a quasistatic picture which focused on the secondary tone's influence on the response of the primary tone and ignored nonlinearity-induced interaction between the responses at the two tones. However, the complex beating patterns observed in Fig.~\ref{fig:Fixed_Detuning_Trajectories} suggest that the interplay between the two drives generates dynamics that the single-tone ansatz [Eq.~\eqref{eq:rotating_quadratures}] cannot capture. To clarify the spectral content of these dynamics, we compute the Fourier transform of the trajectories in Fig.~\ref{fig:Fixed_Detuning_Trajectories}.  Tracking different Fourier components allows the definition of a "stationary state", even though the dynamics appear non-stationary in a single-frequency rotating frame. As shown in Fig.~\ref{fig:fft_trajectories}, the spectrum reveals a significant component at the detuning frequency $\omega=\Delta_{21}$, corresponding to the $\Omega_{2}$ response in the rotating frame. 
To capture this, we extend the single-tone ansatz [Eq.~\eqref{eq:rotating_quadratures}] to a superposition of two rotating frames, treating both driving tones on an equal footing:
\begin{equation}\label{eq:extended_ansatz}
    x(t) = \sum_{i=1,2} X_{i}(T) \cos{(\Omega_{i}t)}-Y_{i}(T) \sin{(\Omega_{i}t)} \, ,
\end{equation}
where $X_{i}$, $Y_{i}$ are the quadratures corresponding to the response at $\Omega_{i}$.
\begin{figure*}
    \includegraphics[]{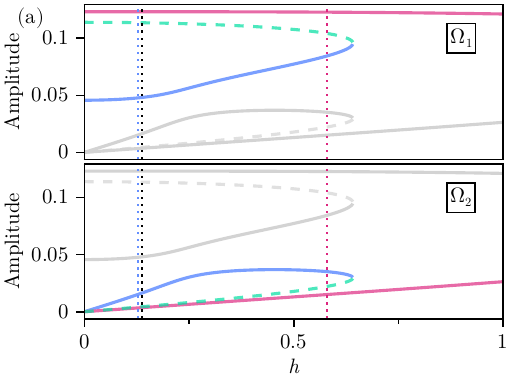}
    \includegraphics[]{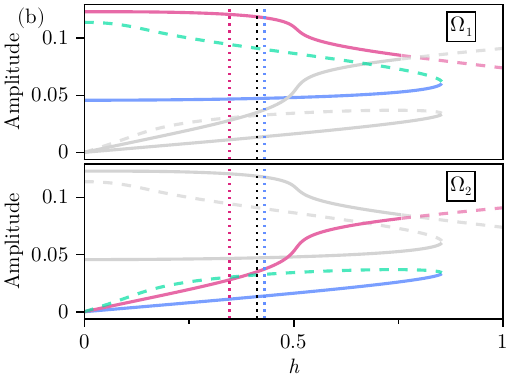}
    \caption{Stationary states obtained from HBM using a multifrequency ansatz [cf.~Eqs.~\eqref{eq:extended_two_tone_HarmonicEqns}] for fixed (a) $\Delta_{21}=0.85\Gamma$ and (b) $\Delta_{21}=-0.85\Gamma$ as a function of $h$. The coloured lines in the top (bottom) panel represent $A_{1}$ at $\Omega_{1}$ ($A_{2}$ at $\Omega_{2}$) and the grey lines represent $A_{2}$ ($A_{1}$). The colours identify the different branches: with respect to $A_{1}$, blue (pink)  refers to the low- (high-) amplitude response at the main drive tone; the corresponding amplitude $A_{2}$ is colored accordingly. The solid lines are stable branches and the dashed lines are saddles. The dotted vertical lines correspond to the (black, blue, and pink) transition boundaries shown in Fig.~\ref{fig:fig4_AvgAmp_PhaseDiagram} for the respective  $\Delta_{21}$. Other parameters are $\Omega_{0}=2\pi$, $\Gamma=6.2\times10^{-4}$, $F_{1}=4.81\times10^{-4}$, and $\Delta_{10}=-9.02\times10^{-4}$.}
    \label{fig:stationary_amp_fixedDetuning}
\end{figure*}

We use the multifrequency ansatz~\eqref{eq:extended_ansatz} and apply the HBM, similar to Section~\ref{sec:OneNLresp}. In the far-detuned regime ($\Delta_{21}\gg \Gamma$), the system's response time $\sim 2\pi/\Gamma$ is much longer than the beating period $2\pi/|\Delta_{21}|$. Thus, we can assume that the quadratures evolve on a timescale $T$ much slower than the driving periods, $T_{i}=2\pi/\Omega_{i}$. This yields four coupled first-order differential equations in $X_{i}$, $Y_{i}$:
\begin{subequations}\label{eq:extended_two_tone_HarmonicEqns}
\begin{align}
\dot{X}_1 &= -\frac{\Gamma}{2} X_1 + \left[ \frac{\tilde{\Delta}^2_{01}}{2\Omega_1} + \frac{3\alpha}{4\Omega_1}(\frac{A_1^2}{2}+A_2^2) \right] Y_1 + \frac{F_1\sin{\theta_1}}{2\Omega_1} \, , \\
\dot{Y}_1 &= -\left[ \frac{\tilde{\Delta}^2_{01}}{2\Omega_1} + \frac{3\alpha}{4\Omega_1}(\frac{A_1^2}{2}+A_2^2) \right] X_1 - \frac{\Gamma}{2} Y_1 + \frac{F_1\cos{\theta_1}}{2\Omega_1}\, , \\
\dot{X}_2 &= -\frac{\Gamma}{2} X_2 + \left[ \frac{\tilde{\Delta}^2_{02}}{2\Omega_2} + \frac{3\alpha}{4\Omega_2}(\frac{A_2^2}{2}+A_1^2) \right] Y_2 + \frac{F_2 \sin{\theta_2}}{2\Omega_2}\, , \\
\dot{Y}_2 &= -\left[ \frac{\tilde{\Delta}^2_{02}}{2\Omega_2} + \frac{3\alpha}{4\Omega_2}(\frac{A_2^2}{2}+A_1^2) \right] X_2 - \frac{\Gamma}{2} Y_2 + \frac{F_2 \cos{\theta_2}}{2\Omega_2}\, ,
\end{align} 
\end{subequations}
where $A_{i}=\sqrt{X_{i}^2+Y_{i}^2}$ is the response amplitude at frequency $\Omega_{i}$, in analogy with Eq.~\eqref{eq:1toneSlowFlowEqDuffing}. 
The nonlinearity introduces self-Kerr terms ($\propto A_i^2$), where the resonance frequency shifts based on the response at the tone itself, and cross-Kerr terms ($\propto A_j^2$), where the response at one tone shifts the effective resonance frequency of the other. These cross-terms capture the non-degenerate four-wave mixing interactions that were averaged out in the single-tone slow-beating approximation.
By balancing harmonics only at $\Omega_1$ and $\Omega_2$ to obtain Eqs.~\eqref{eq:extended_two_tone_HarmonicEqns}, we neglect higher harmonics, e.g., ($3\Omega_{i}$, $5\Omega_{i}$) and mixed tones ($2\Omega_{1}-\Omega_{2}$, $2\Omega_{2}-\Omega_{1}$, ...).

Solving for the stationary states of Eqs.~\eqref{eq:extended_two_tone_HarmonicEqns} presents a significant challenge: the system comprises of four coupled cubic polynomials with a theoretical upper limit of 81 solutions~\cite{bezout1779}. Standard local root-finding algorithms risk missing coexisting stable states in such a complex landscape. To overcome this, we use the \texttt{HarmonicBalance.jl} package~\cite{kosata_harmonicbalancejl_2022}, which relies on homotopy continuation~\cite{breiding_homotopycontinuationjl_2018}. This approach allows us to globally map the solution space, rigorously tracking the number and stability of all stationary states as we sweep the control parameters $h$ and $\Delta_{21}$.

We first study the effect of the relative drive strength $h$ for a fixed detuning $\Delta_{21}$. In Fig.~\ref{fig:stationary_amp_fixedDetuning}, we plot the stationary response amplitudes $A_{1}$ and $A_{2}$ (at frequencies $\Omega_{1}$ and $\Omega_{2}$, respectively) as a function of $h$, assuming the system is initialized in the low-amplitude attractor. For $h=0$, the system reduces to the single-tone case and the amplitude at $\Omega_{2}$ is $0$. For positive detuning [$\Delta_{21}>0$, see Fig.~\ref{fig:stationary_amp_fixedDetuning}(a)], the response at $\Omega_2$ increases steadily with $h$, while the response at $\Omega_1$ remains initially confined to the low-amplitude state (blue branch in the $A_{1}$ panel). However, as the energy injected by the second tone increases with $h>0.25$, the amplitude of this lower branch at $\Omega_{1}$ rises rapidly until the branch destabilizes via a saddle-node bifurcation at $h=0.65$. Beyond this critical threshold, the system transitions to the high-amplitude branch (pink branch in the $A_{1}$ panel) at $\Omega_{1}$, which remains nearly constant, while the response at $\Omega_{2}$ continues to increase with $h$.

The stationary states presented in Fig.~\ref{fig:stationary_amp_fixedDetuning}(a) reveal the mechanism preventing the secondary tone from becoming the dominant spectral component, despite its proximity to the bare resonance. Initially, the system exhibits a linear response regime where $A_2$ grows proportionally to the drive strength $h$, corresponding to the stable dynamics of Region \protect{\circled{1}}. 
However, as the secondary amplitude becomes significant, the cross-Kerr interaction exerts a nonlinear frequency shift that pushes the tone's effective resonance frequency away from $\Omega_2$ (towards lower frequencies for $\alpha < 0$). This interaction creates a scenario reminiscent of an avoided crossing between the responses at $\Omega_{1}$ and $\Omega_{2}$: rather than $A_2$ capturing the input energy and increasing, beyond $h=0.25$ it plateaus and the renormalization of the resonance frequency redistributes power to pump $A_1$, the response at $\Omega_{1}$ (Region \protect{\circled{2}}). Consequently, the nonlinearity ``protects'' the dominance of the primary response $A_{1}$, as the growing secondary tone pushes the effective resonance further towards $\Omega_1$. Ultimately, this cross-Kerr induced pumping drives $A_1$ to a critical threshold where the low-amplitude branch destabilizes via a saddle-node bifurcation, forcing the transition to the high-amplitude attractor observed in Region \protect{\circled{3}}.

For negative detuning [$\Delta_{21}<0$, see Fig.~\ref{fig:stationary_amp_fixedDetuning}(b)], the stationary states display a distinct competition compared with the positive detuning case. The low-amplitude branch at $A_{1}$ and $A_2$ (solid blue line) remains stable over a wide range of $h$, with the primary response $A_1$ dominating. In contrast, the high-amplitude branch at $A_{1}$ and $A_2$ (solid pink line) exhibits a crossover where the secondary response $A_2$ grows rapidly, surpassing $A_1$ near $h \approx 0.5$, shortly before the branch destabilizes. As $h \rightarrow 1$, a parameter window emerges where no stable states exist at the ansatz frequencies.

This behavior correlates directly with the transition thresholds discussed in Fig.~\ref{fig:fig4_AvgAmp_PhaseDiagram}. The low-amplitude branch at $A_{1}$ and $A_{2}$ persists because the secondary tone is far-detuned from the bare resonance; it grows slowly and exerts negligible cross-Kerr influence on the primary mode. Conversely, for the high-amplitude state, the softening nonlinearity pulls the effective resonance frequency down, closer to the negatively detuned secondary tone. This reduces the effective detuning, enhancing the interaction through cross-Kerr terms and causing the branch to exhibit a large $A_{2}$. The crossover occurs in the vicinity of the blue/black transition threshold lines in Fig.~\ref{fig:fig4_AvgAmp_PhaseDiagram}; subsequently, the high-amplitude branch destabilizes, leaving the low-amplitude branch as the only available stable state. 
The absence of any stable states beyond $h=0.775$ indicates that the slow-flow variables $X_i, Y_i$ never settle to a fixed point but instead oscillate. This corresponds to a regime of limit cycle dynamics~\cite{JdelPino_limtcycles_2024}, manifesting physically as persistent, self-sustained amplitude modulations.

\begin{figure}
    \centering
    \includegraphics[]{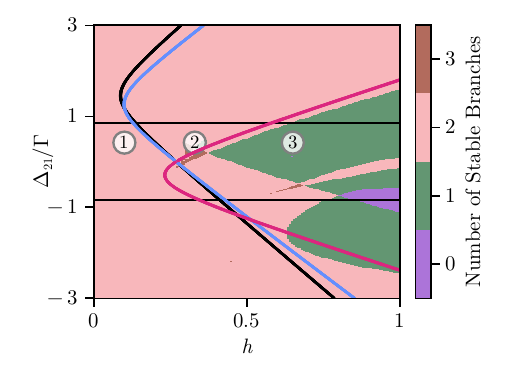}
    \caption{The number of stable solutions plotted against varying $\Delta_{21}$ and $h$ for fixed $\Omega_{0}=2\pi$, $\Gamma=6.2\times10^{-4}$, $F_{1}=4.81\times10^{-4}$, and $\Delta_{10}=-9.02\times10^{-4}$. The black lines at $\Delta_{21}=\pm0.85\Gamma$ reflect the cuts shown in Fig.~\ref{fig:stationary_amp_fixedDetuning}. \protect\circled{1} - \protect \circled{3} represent the fixed parameters used for the trajectories in Figs.~\ref{fig:Fixed_Detuning_Trajectories} and~\ref{fig:fft_trajectories}. The colored transition thresholds shown in Fig.~\ref{fig:fig4_AvgAmp_PhaseDiagram} have been superposed for comparison. A region with zero stable fixed points (purple) emerges, indicative of limit cycle dynamics.}
    \label{fig:phasediagram_HB_2Tones}
\end{figure}

To map the global stability landscape, we repeat the process above as a function of $h$ and $\Delta_{21}$, counting the number of \textit{stable} branches, see Fig.~\ref{fig:phasediagram_HB_2Tones}. The black horizontal lines ($\Delta_{21}\sim\Gamma$) represent the limit of spectral resolution of the tones; the two-tone HBM is robust for well-resolved tones (i.e, $|\Delta_{21}|>\Gamma$). For positive detunings ($\Delta_{21} > 0$), the numerically determined boundary where the system transitions from bistability (two stable branches, pink region) to monostability (one stable branch, green region) aligns remarkably well with the transition threshold derived from the high-amplitude renormalized filtered effective drive model (solid pink line). 
This confirms that, in the positively detuned regime ($\Delta_{21}>0$), the transition is fundamentally governed by a strong enough drive overcoming the resonance frequency shift induced by the target state itself.

In contrast, the negative detuning region ($\Delta_{21} < 0$) reveals the breakdown of the stationary two-tone ansatz, consistent with the complex competition observed in Fig.~\ref{fig:stationary_amp_fixedDetuning}(b). 
Here, the bifurcation boundaries diverge from the effective drive transition thresholds, indicating that the cross-Kerr interaction with the negatively detuned secondary tone induces parametric instabilities not captured by the simple resonance frequency shift model. We also observe that a distinct region emerges at high $h$ where the number of stable stationary branches drops to zero (purple region). This absence of stable fixed points provides strong evidence for the formation of limit cycles, confirming that the dynamical phase transitions in the negatively detuned regime involve time-dependent non-stationary states rather than simple inter-attractor switching.

The multifrequency ansatz [cf.~Eq.~\eqref{eq:extended_ansatz}] and the resulting equations of motion [cf.~Eq.~\eqref{eq:extended_two_tone_HarmonicEqns}] successfully capture the cross-Kerr interaction (non-degenerate four-wave mixing) between the system's responses to the two drives, providing physical insight beyond the linear and slow-beating approximations. The accuracy of this approach is confirmed in Fig.~\ref{fig:fft_trajectories}, where the HBM predictions at parameters \circled{1}-\circled{3} align well with the distinct $\Omega_1$ and $\Omega_2$ components extracted from the Fourier spectra of the time-domain trajectories. However, the method relies on the assumption of slowly varying envelopes, which is strictly valid only when the tones are spectrally resolved ($|\Delta_{21}| \gtrsim \Gamma$). As the system approaches the slow-modulation limit ($\Delta_{21} \to 0$), this timescale separation breaks down. This leads to deviations between the multiple sharp bifurcation thresholds predicted by the HBM and the transitions observed in the phase diagram (cf.~Fig.~\ref{fig:fig4_AvgAmp_PhaseDiagram}), which are further broadened by the coarse-graining inherent to the time-averaged order parameter $\bar{A}$. Finally, remaining discrepancies—particularly in the negative detuning regime—suggest that higher-order intermodulation products (e.g., $2\Omega_1 - \Omega_2$) neglected by the two-tone truncation may play a non-negligible role in the dynamics.

\section{\label{sec:Conclusion/Outlook}Conclusion/Outlook}

In this work, we have developed a comprehensive framework to describe dynamical phase transitions in a Duffing resonator under bichromatic driving. By employing two complementary analytical approaches—an effective nonlinear response theory and an extended multifrequency ansatz—we have moved beyond phenomenological observation to uncover the distinct mechanisms governing the transitions in this system. Our analysis reveals that even a weak secondary tone can profoundly reshape the dynamics, not merely as a perturbation, but by fundamentally altering the stability landscape.

Our effective response theory provides a predictive tool for the slow-beating regime, where the secondary tone acts as a modulation that pushes the system past bifurcation points. Crucially, we identify that the onset of inter-attractor switching is not governed solely by the initial state, but by the resonance properties of the \textit{target} state. Complementing this, our extended two-tone ansatz, valid in the resolved-sideband regime ($|\Delta_{21}| \gtrsim \Gamma$), reveals the microscopic mechanism behind these shifts. It reveals how cross-Kerr interactions (non-degenerate four-wave mixing) can either suppress switching via an avoided-crossing-like effect or induce instabilities leading to limit cycles, particularly in the negatively-detuned ($\Delta_{21}<0$) regime.

In contrast to earlier studies that surveyed specific parameters for sensing applications~\cite{houri_pulse-width_2019} or focused on cascades to chaos~\cite{houri_generic_2020}, we provided a rigorous mapping of the instability thresholds across a wide parameter space. This allowed us to explain the marked asymmetry between negative ($\Delta_{21}<0$) and positive ($\Delta_{21}>0$) detuning observed in experiments. Our results thereby offer a qualitative yet predictive framework to categorize dynamical phase transitions across nanomechanical~\cite{michaeli_optically_2025}, optical~\cite{bloch_param_coupled_2020}, and superconducting resonator~\cite{fanisani_level_2021, blais_cqed_2021} platforms.

Looking ahead, this framework lays the foundation for systematic extensions. The multifrequency approach can be expanded to include higher-order intermodulation products, analytically clarifying the route to chaos in multi-tone driven systems. Furthermore, our stability maps provide a basis for investigating how stochastic or quantum fluctuations mediate transitions near these newly identified thresholds~\cite{marthaler_switching_2006}. Finally, in the context of strongly nonlinear Kerr resonators, these insights into controlled state-switching offer new protocols for quantum control and qubit operation~\cite{valery_twotoneflux_2022, valentin_bichromsemicondqubit_2024}, establishing multi-tone driving as a versatile resource for engineering dynamical phases.

\section*{Acknowledgements}
We acknowledge funding from the Deutsche Forschungsgemeinschaft (DFG) through project numbers 449653034, 521530974, and 545605411, as well as via SFB 1432 (project number 425217212). We also acknowledge support from the Swiss National Science Foundation (SNSF) through the Sinergia Grant No. CRSII5\_206008/1. J.d.P. acknowledges funding from the Ramón y Cajal program (RYC2023-043827-I), funded by MICIU/AEI (10.13039/501100011033) and FSE+, and through the “María de Maeztu” Programme for Units of Excellence in R\&D (CEX2023-001316-M).
    
\section*{Appendices}
\setcounter{section}{0}
\renewcommand{\thesection}{\Alph{section}}
\appendix
\input{Appendix}

\bibliography{bibliography}
\end{document}

%% file: Appendix.tex
\section{Derivation of the linear modulated power}
\label{sec:derivation_linearpower_modulated}
In this appendix, we provide a detailed derivation of the coupled equations of motion for the quadratures $X(t)$ and $Y(t)$ of a two-tone driven linear resonator, which are presented in Sec.~III. Our starting point is the equation of motion for a linear damped harmonic oscillator, which is obtained by setting the nonlinearity $\alpha=0$ in Eq.~\eqref{eq:DuffingEOMTwoTone}:
\begin{equation}\label{eq:appendixeq1}
\ddot{x} + \Gamma \dot{x} + \Omega_0^2 x = \text{Re}\left[\sum_{i=1,2} F_i e^{i(\Omega_i t + \theta_i)}\right]\,.  
\end{equation}

We reformulate the problem in a frame co-rotating with the primary drive at frequency $\Omega_1$. To this end, we employ the transformation for the resonator displacement $x(t)$ given in Eq.~\eqref{eq:rotating_quadratures}:
\begin{equation} \label{eq:appendixeq2}
x(t) = X(t)\cos(\Omega_1 t) - Y(t)\sin(\Omega_1 t)\,.
\end{equation}
with first and second time derivatives given by
\begin{align} \label{eq:appendixeq3}
\dot{x}(t) =& [\dot{X}(t) - \Omega_1 Y(t)]\cos(\Omega_1 t) \nonumber \\
&- [\dot{Y}(t) + \Omega_1 X(t)]\sin(\Omega_1 t)\,,\\
\ddot{x}(t) = &[\ddot{X}(t) - 2\Omega_1 \dot{Y}(t) - \Omega_1^2 X(t)]\cos(\Omega_1 t) \nonumber \\
&- [\ddot{Y}(t) + 2\Omega_1 \dot{X}(t) - \Omega_1^2 Y(t)]\sin(\Omega_1 t)\,. 
\end{align}
\begin{figure}[t!]
    \centering
    \includegraphics[]{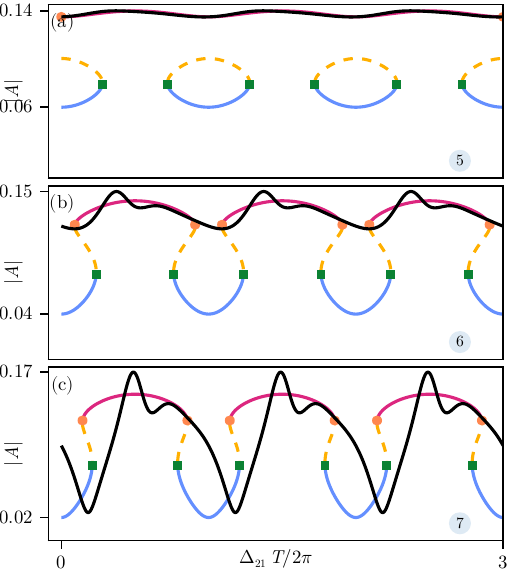}
    \caption{The trajectories as in Fig.~\ref{fig:Fixed_Detuning_Trajectories} for cases \protect \circled{5}-\protect\circled{7} in Table~\ref{tab:transition_regimes} for fixed detuning of the secondary tone, $\Delta_{21}=0.45\Gamma$, and relative strength (a) $h=0.07$, (b) $h=0.275$ and (c) $h=0.55$. The parameters used are $\Omega_{0}=2\pi$,$\Delta_{10}=-9.02\times10^{-4}$, $\Gamma=6.2\times10^{-4}$, and $F_{1}=6.01\times10^{-4}$.}
    \label{fig:nearlb_fixed_trajs}
\end{figure}
We decompose the second tone with respect to the first, using the detuning $\Delta_{21} = \Omega_2 - \Omega_1$ and set the drive phases to zero ($\theta_1 = \theta_2 = 0$), such that the total drive is $F(t) = F_1 \cos(\Omega_1 t) + F_2 \cos(\Omega_2 t)$. We express the second tone with reference to the rotating frame frequency $\Omega_1$
\begin{align} \label{eq:appendixeq5}
F(t) &= F_1 \cos(\Omega_1 t) + F_2 \cos(\Omega_1 t + \Delta_{21} t) \nonumber \\
&= F_X(t)\cos(\Omega_1 t) + F_Y(t) \sin(\Omega_1 t)\, .
\end{align}
with $F_X(t)=F_1(1 + h \cos(\Delta_{21} t))$ and $F_Y(t)=-hF_1 \sin(\Delta_{21} t)$ the time-dependent drive amplitudes into in-phase and quadrature components, for modulation amplitude $h=F_2/F_1$.
We proceed by substituting the expressions for $x(t)$ and its derivatives [Eqs. \eqref{eq:appendixeq2}-\eqref{eq:appendixeq3}] and for the force [Eq. \eqref{eq:appendixeq5}] into the equation of motion [Eq. \eqref{eq:appendixeq1}]. We then apply the method of harmonic balancing by collecting all terms proportional to $\cos(\Omega_1 t)$ and $\sin(\Omega_1 t)$ and equating them respectively, to $F_X(t)$ and $F_Y(t)$. We obtain:
\begin{equation}\label{eq:appendixeq6_7}
\begin{pmatrix}
\partial_{tt} -\Omega_1^2+\Omega_0^2+\Gamma\partial_t & -2\Omega_1\partial_t -\Gamma\Omega_1 \\
2\Omega_1\partial_t +\Gamma\Omega_1 & \partial_{tt} -\Omega_1^2+\Omega_0^2+\Gamma\partial_t\end{pmatrix}
\vec{R}
=
 \vec{F}_1(t),
\end{equation}
where $\partial_t\square\equiv \dot{\square}$ denotes the first time derivative,
$\partial_{tt}\square\equiv\ddot{\square}$ the second time derivative, $\vec{R}=(X,Y)^T$ and $\vec{F}_1(t)=(F_X(t),F_Y(t))^T$. 
These equations fully describe the dynamics of the linear system in the rotating frame, accounting for the beating between the two drive tones.

The linear system in Eqs.~\eqref{eq:appendixeq6_7} can be solved exactly by Fourier transforming the equations of motion, which converts them into coupled algebraic relations for the spectral amplitudes $\tilde{X}(\omega)$ and $\tilde{Y}(\omega)$. In this domain, each time derivative becomes $(i\omega)^n$, and the drives $F_X(t),F_Y(t)$ appear as Dirac delta peaks at $\omega=0$ and $\omega=\pm\Delta_{21}$.
Solving the algebraic system for $\tilde{X}(\omega)$ and $\tilde{Y}(\omega)$ and transforming back to time gives the quadratures $X(t)$ and $Y(t)$. The instantaneous power in the rotating frame is $A^2_{\raisebox{-0.3ex}{$\scriptscriptstyle \Omega_0$}}(t, h)=X^2+Y^2$, leading to the modulated response at $\Omega_1$ shown in Eq.~\eqref{eq:linearPoweratPumptimedependence}.
To locate the extrema of the modulated power, we set $\partial_t A^2_{\raisebox{-0.3ex}{$\scriptscriptstyle \Omega_0$}}(t, h)=0$. Solving for the critical amplitudes gives the cycle’s maximum and minimum power, cf. Eq.~\eqref{eq:LinearPowerMaxMin}.
This expression is central to the model developed in Sec.~\ref{sec:Phase Transition Boundaries} for determining the phase transition boundaries.

\section{Exploration of more dynamical cases}
\label{sec:additional_cases}
\begin{figure}[t!]
    \centering
    \includegraphics[]{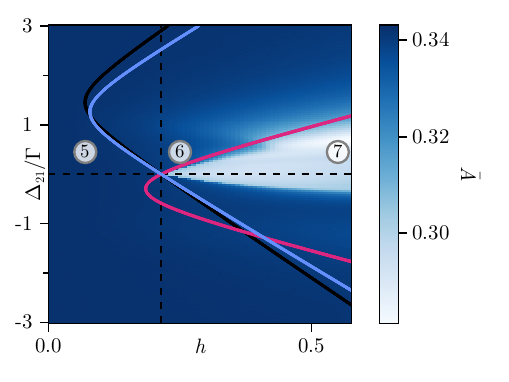}
    \caption{The average amplitude phase diagram with threshold lines [cf.~Fig.~\ref{fig:fig4_AvgAmp_PhaseDiagram}(c)], for an initial point starting from the higher branch, for $F_1 \sim F_{lb}$. The parameters used are $\Omega_{0}=2\pi$, $\Delta_{10}=-9.02\times10^{-4}$, $\Gamma=6.2\times10^{-4}$, and $F_{1}=6.01\times10^{-4}$.}
    \label{fig:nearlb_starthi}
\end{figure}

\begin{figure*}[t!]

    \includegraphics[]{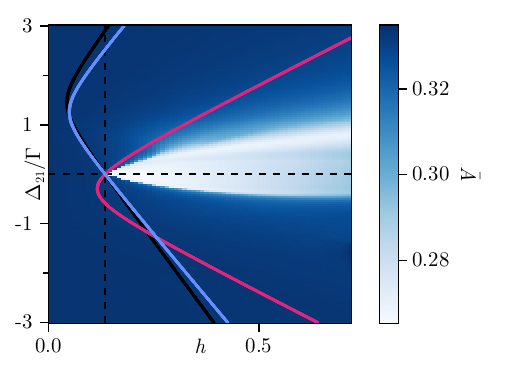}\hfill
    \includegraphics[]{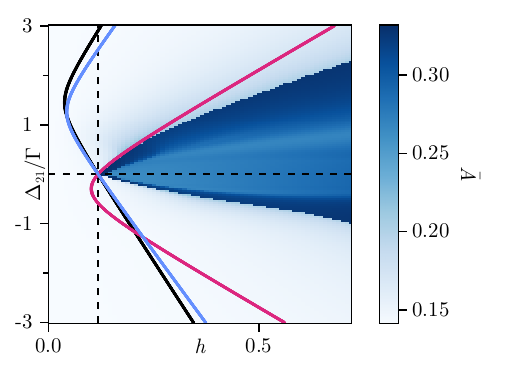}
    
    \caption{The average amplitude phase diagram with threshold lines [cf.~Fig.~\ref{fig:fig4_AvgAmp_PhaseDiagram}(c)] for initial point in the middle of the bistability region. (a) The initial point starts from the higher branch. (b) The initial point starts from the lower branch. The parameters used are $\Omega_{0}=2\pi$,$\Delta_{10}=-9.02\times10^{-4}$, $\Gamma=6.2\times10^{-4}$, and $F_{1}=5.46\times10^{-4}$.}
    \label{fig:mid_start_combined}
\end{figure*}

Our model's framework extends to other dynamical regimes. For instance, initializing the system on the high-amplitude branch captures the complementary transitions to Fig.~\ref{fig:fig4_AvgAmp_PhaseDiagram}(c), given by cases \circled{5}-\circled{8} in Table~\ref{tab:transition_regimes} and Fig.~\ref{fig:nearlb_fixed_trajs}.
The comparison of ${F_\mathrm{eff}}_{\raisebox{-0.3ex}{$\scriptscriptstyle \Omega, -$}}$ with $F_{hb}$ gives the threshold that correctly predicts the jumps down to the low branch in the corresponding phase diagram, see Fig.~\ref{fig:nearlb_starthi}.

Conversely, when the system is initialized in the center of the bistability region, the model's quantitative agreement deviates, see Fig.~\ref{fig:mid_start_combined}. This is an expected outcome. Our model is based on a linear-response approximation, which is most accurate on the quasi-linear part of the branches [cf. Fig.~\ref{fig:Fmod_SharkShift}(c)], i.e., far from their respective bifurcations. In the middle of the bistable region, the branches begin to deviate from the quasi-linear behavior as they approach their respective bifurcations. Here, the system's response amplitude can no longer be approximated using the two-tone driven linear resonator's response [cf. Eq.~\ref{eq:linearPoweratPumptimedependence}, leading to deviations between the predicted threshold and the simulations.

While asymptotic expansions can analyze such systems in the slow-modulation limit ($\Delta_{21}\ll \Gamma$) by leveraging timescale separation~\cite{desroches_mixedmode_2012, zhuDelayedSNB2015, desroches_bursting_minmodel_2019}, our model's strength lies in providing a qualitative and intuitive framework even when this separation is lost (i.e., as $\Delta_{21}$ approaches $\Gamma$). It successfully pinpoints the underlying mechanism—the competition between the drives due to the state-dependent resonance frequency—that governs the rich inter-attractor dynamics, offering a predictive tool where other analytical approaches are no longer applicable.

\section{Experimental platform}
\label{sec:experimental_setup}
\begin{figure}[t!]
    \centering
    \includegraphics[width=\columnwidth]{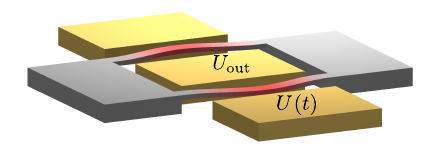}
    \caption{Schematic representation of double-ended tuning fork device (adapted from \cite{catalini_slow_2025}). The gray area represents the double-ended tuning fork with the vibrating part highlighted in red. The two outer gold pads represent the electrodes where we apply the driving voltage $U(t)$, the central one represents the one from which we readout the signal. The electrode for the bias voltage is not represented.}
    \label{fig:figapp1}
\end{figure}
Here we provide details on the experimental setup and the measurement protocol used to generate Fig. \ref{fig:fig4_AvgAmp_PhaseDiagram}(a). The experimental setup consists of a microelectromechanical system (MEMS) shaped as a double-ended tuning fork with branches $200\,\si{\mu m}$ long and $6\,\si{\mu m}$ thick. The device is capacitively coupled to gold electrodes fabricated next to it, which we can use to apply a driving $U(t)$ and bias voltage $U_\mathrm{b}$, and to readout the signal $U_{out}$ containing the information about the resonator motion, see Fig. \ref{fig:figapp1}. The bias voltage is used to tune the resonance frequency and the Duffing parameter. The device is fabricated by Prof. Kenny's group in Stanford \cite{Agarwal_2008} and is made out of highly-doped single crystal silicon with a p-type (Boron) concentration of $1.5 \times 10^{20}\, \si{\cm}^{-3}$. The resonator is maintained at a pressure of $10^{-1}\,\si{mbar}$ through an epi-seal process \cite{Yang2016}.

The resonator vibration can be decomposed over a set of mechanical modes, which, in the limit of weak drive, can be individually described with the equation of motion of a simple damped harmonic oscillator:

\begin{equation}
\ddot{x}+\Omega_n^2 x+\Gamma_n\dot{x}=F_\mathrm{dr}(t),
\end{equation}
where $\Omega_n$ and $\Gamma_n$ are the resonance frequency and the damping rate of the $n-$th mode, respectively, and $F_\mathrm{dr}(t)=\text{K} U(t)$ is the applied drive in units of $\si{V}\,\si{s}^{-2}$. $\text{K}$ is a conversion factor expressed in units of $\si{s}^{-2}$. When the strength of the external driving force increases, the resonator behavior starts to deviate from the standard damped harmonic oscillator. As we enter such large amplitude oscillation limit, to describe the system response to an external driving force we need to use the Duffing equation:

\begin{equation}
\ddot{x}+\Omega_n^2 x+\Gamma_n\dot{x}+\alpha_n x^3=\text{Re}\left[F_\mathrm{dr}e^{i(\Omega_\mathrm{dr}t+\theta)}\right],
\end{equation}
where $\alpha$ is the Duffing parameter in units of $\si{V}^{-2}\,\si{s}^-2$. 

In this work, we only focus on the lowest mechanical mode of the resonator. The mechanical parameters for this particular device have been characterized in a previous work \cite{catalini_slow_2025} by measuring the amplitude ($A$) and phase ($\phi$) response to an external drive in both the weak and strong drive limit. Using a lock-in amplifier, we apply an external driving force $\Omega_{dr}$ to the system, varying its value across the resonance $\Omega_0$. With the same instrument, we simultaneously measure $A$ and $\phi$ in the frame rotating at $\Omega_{dr}$. The measured mechanical parameters are $\Omega_0/2\pi=1.11\,\si{MHz}$, $\Gamma/2\pi= 108\,\si{Hz}$, $\alpha=-1.89\,\si{V}^{-2}\,\mathrm{s}^{-2}$ and $\text{K}\approx 1\times 10^7 s^{-2}$.

\begin{figure}[t!]
    \centering
    \includegraphics[width=\columnwidth]{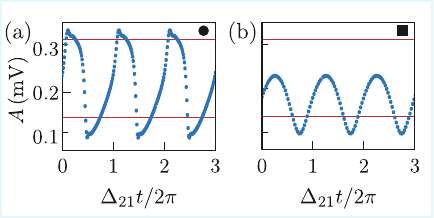}
    \caption{Experimental phase diagram. Example of a measured $A(t)$ for (a) $\Delta_{21}<\Gamma$ and (b) $\Delta_{21}\approx \Gamma$ collected with the same $F_1$, $\Delta_{10}$ and $h$. The upper (lower) red line represents the upper (lower) stable solution in the absence of the second drive (adapted from \cite{catalini_slow_2025}).}
    \label{fig:figapp2}
\end{figure}

To generate the phase diagram in Fig.~\ref{fig:fig4_AvgAmp_PhaseDiagram}(a), we measure the resonator amplitude $A$ as a function of time when driven with two different tones $F_1e^{i(\Omega_1t)}$ and $F_2e^{i(\Omega_2t))}$. We use the lock-in amplifier to apply the two driving tones and readout the amplitude in the frame rotating at $\Omega_1$. After fixing $F_1/\text{K}=140\mathrm{mV}$ and $\Delta_{10}/2\pi\approx -158\,{Hz}$, we systematically probe the response of the resonator while varying $\Delta_{21}=\Omega_2-\Omega_1$ and $h$. For each combination, we extract the average value $\bar{A}$ for an interval of $5\,\si{s}$ making sure that for each combination the system is initialized in the lower stable solution of the Duffing curve before turning on the second drive. An example of measured $A(t)$ in the two limits of $\Delta_{21}<\Gamma$ and $\Delta_{21}\approx\Gamma$ are shown in Fig.~\ref{fig:figapp2}. 

%% file: main.bbl
\begin{thebibliography}{63}
\bibitem [{Fou()}]{Fourier}%
  \BibitemOpen
  \href@noop {} {}\bibinfo {note} {Note that, the Fourier transform assumes the signal is defined for all time, effectively requiring an infinitely long observation.}\BibitemShut {Stop}%
\end{thebibliography}
